
\input harvmac
\input amssym.def
\input amssym
\input epsf.tex
\baselineskip 14pt
\magnification\magstep1
\parskip 6pt
\newdimen\itemindent \itemindent=32pt
\def\textindent#1{\parindent=\itemindent\let\par=\resetpar%
\indent\llap{#1\enspace}\ignorespaces}

\let\oldpar=\par
\def\resetpar{\oldpar\parindent=20pt\let\par=\oldpar}

\font\ninerm=cmr9 \font\ninesy=cmsy9
\font\eightrm=cmr8 \font\sixrm=cmr6
\font\eighti=cmmi8 \font\sixi=cmmi6
\font\eightsy=cmsy8 \font\sixsy=cmsy6
\font\eightbf=cmbx8 \font\sixbf=cmbx6
\font\eightit=cmti8
\def\eightpoint{\def\rm{\fam0\eightrm}
  \textfont0=\eightrm \scriptfont0=\sixrm \scriptscriptfont0=\fiverm
  \textfont1=\eighti  \scriptfont1=\sixi  \scriptscriptfont1=\fivei
  \textfont2=\eightsy \scriptfont2=\sixsy \scriptscriptfont2=\fivesy
  \textfont3=\tenex   \scriptfont3=\tenex \scriptscriptfont3=\tenex
  \textfont\itfam=\eightit  \def\it{\fam\itfam\eightit}%
  \textfont\bffam=\eightbf  \scriptfont\bffam=\sixbf
  \scriptscriptfont\bffam=\fivebf  \def\bf{\fam\bffam\eightbf}%
  \normalbaselineskip=9pt
  \setbox\strutbox=\hbox{\vrule height7pt depth2pt width0pt}%
  \let\big=\eightbig  \normalbaselines\rm}
\catcode`@=11 %
\def\eightbig#1{{\hbox{$\textfont0=\ninerm\textfont2=\ninesy
  \left#1\vbox to6.5pt{}\right.\n@@space$}}}
\def\vfootnote#1{\insert\footins\bgroup\eightpoint
  \interlinepenalty=\interfootnotelinepenalty
  \splittopskip=\ht\strutbox %
  \splitmaxdepth=\dp\strutbox %
  \leftskip=0pt \rightskip=0pt \spaceskip=0pt \xspaceskip=0pt
  \textindent{#1}\footstrut\futurelet\next\fo@t}
\catcode`@=12 %
\def \d{{\rm d}}
\def \de{\delta}
\def \De{\Delta}

\def \eps{\epsilon}

\def \pr{\partial}
\def \d{{\rm d}}

\def \l{\big \langle}
\def \r{\big \rangle}

\def \half{{\textstyle {1 \over 2}}}

\def \quar{{\textstyle {1 \over 4}}}
\def \ts{\textstyle}
\def \A{{\cal A}}
\def \B{{\cal B}}
\def \C{{\cal C}}
\def \D{{\cal D}}

\def \F{{\cal F}}
\def \G{{\cal G}}
\def \H{{\cal H}}
\def \I{{\cal I}}
\def \J{{\cal J}}
\def \K{{\cal K}}

\def \M{{\cal M}}
\def \N{{\cal N}}

\def \S{{\cal S}}

\def \bx{\bar x}

\def \vphi{{\varphi}}
\def \tF{{\tilde{\cal F}}}
\def \hf{{\hat f}}
\def \ta{t}
\def \oD{\overline{D}}
\def \Li{\mathop{\rm Li}\nolimits}
\font \bigbf=cmbx10 scaled \magstep1


\lref\DO{F.A. Dolan and H. Osborn, Conformal Four Point Functions and the
Operator Product Expansion, Nucl. Phys. B599 (2001) 459,  hep-th/0011040.}
\lref\SCFT{F.A. Dolan and H. Osborn, Superconformal Symmetry, Correlation 
Functions and the Operator Product Expansion, Nucl. Phys. B629 (2002) 3,
hep-th/0112251.}
\lref\NO{M. Nirschl and H. Osborn, Superconformal Ward Identities and
their Solution, Nucl. Phys. B711 (2005) 409, hep-th/0407060.}
\lref\Short{F.A. Dolan and H. Osborn, {On short and semi-short
representations for four-dimensional superconformal symmetry}, Ann. Phys.
307 (2003) 41, hep-th/0209056.}

\lref\Lang{K. Lang and W. R\"uhl, The Critical $O(N)$ Sigma Model at Dimension
$2<d<4$ and order $1/N^2$: Operator Product Expansions and Renomalization,
Nucl. Phys. {B377} (1992) 371\semi
K. Lang and W. R\"uhl, The Critical $O(N)$ Sigma Model at Dimension
$2<d<4$ and order $1/N^2$: Fusion Coefficients and Anomalous Dimensions,
Nucl. Phys. {B400} (1993) 597\semi
K. Lang and W. R\"uhl, The Critical $O(N)$ Sigma Model at Dimension
$2<d<4$ and order $1/N^2$: A List of Quasiprimary Fields, Nucl. Phys. {B402} 
(1993) 573\semi
K. Lang and W. R\"uhl, The Critical $O(N)$ Sigma Model at Dimension
$2<d<4$ and order $1/N^2$: The Degeneracy of Quasiprimary Fields and its
Resolution, Zeit. f. Physik C61 (1994) 495\semi
K. Lang and W. R\"uhl, The Critical $O(N)$ Sigma Model at Dimension
$2<d<4$ and order $1/N^2$: Hardy-Ramanujan Distribution of Quasiprimary Fields 
and a Collective Fusion Approach, Zeit. f. Physik C63 (1994) 531, 
hep-th/9401116.}

\lref\Cardy{J. Cardy, Anisotropic Corrections to Correlation Functions
in Finite Size Systems, Nucl. Phys. B290 (1987) 355.}
\lref\hughone{H. Osborn and A. Petkou, {Implications of Conformal
Invariance for Quantum Field Theories for General Dimensions},
Ann. Phys. (N.Y.) {231} (1994) 311, hep-th/9307010.}
\lref\Pet{A. Petkou, Conserved Currents, Consistency Relations and Operator
Product Expansions in the Conformally Invariant $O(N)$ Vector Model,
Annals Phys. 249 (1996) 180, hep-th/9410093.}

\lref\Howe{P.J. Heslop and P.S. Howe, Four-point functions in $N=4$ SYM,
JHEP 0301 (2003) 043, hep-th/0211252.}

\lref\Heslop{P.J. Heslop, private communication.}

\lref\Nirschl{M. Nirschl, University of Cambridge PhD thesis, 2005.}

\lref\ASok{F.A. Dolan, L. Gallot and E. Sokatchev, On Four-point Functions
of $\half$-BPS operators in General Dimensions, JHEP 0409 (2004) 056,
hep-th/0405180.}

\lref\Hokone{E. D'Hoker, S.D. Mathur, A. Matusis and L. Rastelli, {The
Operator Product Expansion of $N=4$ SYM and the 4-point Functions of
Supergravity}, Nucl. Phys. B589 (2000) 38, hep-th/9911222.}

\lref\Freed{E. D'Hoker, D.Z. Freedman, S.D. Mathur, A. Matusis and L. Rastelli, 
{Graviton Exchange and Complete 4-point Functions in the AdS/CFT Correspondence},
Nucl. Phys. B562 (1999) 353, hep-th/9903196.}

\lref\Arut{G. Arutyunov and S. Frolov, {Four-point Functions of Lowest
Weight CPOs in $\N=4$ SYM${}_4$ in Supergravity Approximation},
Phys. Rev. D62 (2000) 064016, hep-th/0002170.}

\lref\ADHS{G. Arutyunov, F.A. Dolan, H. Osborn and E. Sokatchev,
Correlation Functions and Massive Kaluza-Klein Modes in the AdS/CFT
Correspondence, Nucl. Phys. B665 (2003) 273, hep-th/0212116.}
\lref\Degen{G. Arutyunov and E. Sokatchev, On a Large N Degeneracy
in $\N=4$ SYM and the AdS/CFT Correspondence, Nucl. Phys. B663 (2003) 163,
hep-th/0301058.}

\lref\OPEN{G. Arutyunov, S. Frolov and A.C. Petkou, {Operator Product
Expansion of the Lowest Weight CPOs in $\N=4$ SYM${}_4$ at Strong Coupling},
Nucl. Phys. B586 (2000) 547, hep-th/0005182;
(E) Nucl. Phys. B609 (2001) 539.}
\lref\OPEW{G. Arutyunov, S. Frolov and A.C. Petkou, {Perturbative and
instanton corrections to the OPE of CPOs in $\N=4$ SYM${}_4$}, Nucl.
Phys. B602 (2001) 238, hep-th/0010137; (E) Nucl. Phys. B609 (2001) 540.}
\lref\Edent{B. Eden, A.C. Petkou, C. Schubert and E. Sokatchev, {Partial
non-renormalisation of the stress-tensor four-point function in $N=4$
SYM and AdS/CFT}, Nucl. Phys. B607 (2001) 191, hep-th/0009106.}
\lref\Except{G. Arutyunov, B. Eden, A.C. Petkou and E. Sokatchev,
{Exceptional non-renorm-alization properties and OPE analysis of
chiral four-point functions in $\N=4$ SYM${}_4$},
Nucl. Phys. B620 (2002) 380, hep-th/0103230.}
\lref\Bia{M. Bianchi, B. Eden, G. Rossi and Y.S. Stanev, On Operator
Mixing in $\N=4$ SYM, Nucl. Phys. B646 (2002) 69, hep-th/0205321.}
\lref\Biamix{M. Bianchi, G. Rossi and Y.S. Stanev, Surprises from the
Resolution of Operator Mixing in $\N=4$ SYM, Nucl. Phys. B685 (2004) 65, 
hep-th/0312228.}

\lref\pert{F. Gonzalez-Rey, I. Park and K. Schalm, {A note on four-point
functions of conformal operators in $N=4$ Super-Yang Mills},
Phys. Lett. B448 (1999) 37, hep-th/9811155\semi
B. Eden, P.S. Howe, C. Schubert, E. Sokatchev and P.C. West,
{Four-point functions in $N=4$ supersymmetric Yang-Mills theory at
two loops}, Nucl. Phys. B557 (1999) 355, hep-th/9811172; {Simplifications
of four-point functions in $N=4$ supersymmetric Yang-Mills theory at
two loops}, Phys. Lett. B466 (1999) 20, hep-th/9906051\semi
M. Bianchi, S. Kovacs, G. Rossi and Y.S. Stanev, {On the
logarithmic behaviour in $\N=4$ SYM theory}, JHEP 9908 (1999) 020,
hep-th/9906188\semi
B. Eden, C. Schubert and E. Sokatchev, Three loop four point
correlator in N=4 SYM, Phys. Lett. B482 (2000) 309, hep-th/0003096.}
\lref\Kon{M. Bianchi, S. Kovacs, G. Rossi and Y.S. Stanev, Anomalous dimensions
in $\N$=4 SYM theory at order $g^4$, Nucl. Phys. B584 (2000) 216,
hep-th/0003203.}
\lref\hpert{G. Arutyunov, S. Penati, A. Santambrogio and E.Sokatchev,
Four-point correlators of BPS operators in $\N=4$ SYM at order $g^4$,
Nucl. Phys. B670 (2003) 103, hep-th/0305060.}

\lref\mixing{G. Arutyunov, S. Penati, A.C. Petkou, A. Santambrogio and 
E.Sokatchev, Non-protected operators in $\N=4$ SYM and multiparticle
states in AdS${}_5$ SUGRA, Nucl. Phys. B643 (2002) 49, hep-th/0206020.}

\lref\HMR{L. Hoffmann, L. Mesref and W. R\"uhl, {Conformal partial wave
analysis of AdS amplitudes for dilaton-axion four-point functions},
Nucl. Phys. B608 (2001) 177, hep-th/0012153\semi
T. Leonhardt, A. Meziane and W. R\"uhl, Fractional BPS Multi-Trace
Fields of $\N=4$ SYM${}_4$ from AdS/CFT, Phys. Lett. B552 (2003) 87,
hep-th/0209184.}

\lref\UD{N.I. Ussyukina and A.I. Davydychev, {An approach to the evaluation
of three- and four- point ladder diagrams}, Phys. Lett. B298 (1993) 363\semi
N.I. Ussyukina and A.I. Davydychev, Exact results for three- and four- point 
ladder diagrams with an arbitrary number of rungs, Phys. Lett. B305
(1993) 136.}
\lref\Isa{A.P. Isaev, {Multi-loop Feynman Integrals and Conformal
Quantum Mechanics}, Nucl. Phys. B662 (2003) 461, hep-th/0303056.}

\lref\lipatov{
A.V. Kotikov, L.N. Lipatov, A.I. Onishchenko and V.N. Velizhanin,
{Three-loop Universal Anomalous Dimension of the Wilson Operators
in $N = 4$ SUSY Yang-Mills model},
Phys. Lett. B595 (2004) 521, hep-th/0404092.}
\lref\belitsky{
A.V. Belitsky, S.E. Derkachov, G.P. Korchemsky and A.N. Manashov,
{Superconformal Operators in $N = 4$ Super-Yang-Mills Theory},
hep-th/0311104.}
\lref\OST{K. Okuyama and L-S. Tseng, Three-Point Functions in $\N=4$ SYM 
Theory at One-Loop, hep-th/0404190.}

\lref\Gub{S.S. Gubser, I.R. Klebanov and A.M. Polyakov, A semi-classical 
limit of the gauge/string correspondence, Nucl. Phys. B636 (2002) 99,
hep-th/0204051.}

{\nopagenumbers
\rightline{DAMTP/04-152}
\rightline{hep-th/0412335}
\vskip 1.5truecm
\centerline {\bigbf Conformal Partial Wave Expansions for}
\vskip 4pt
\centerline {\bigbf $\N=4$ Chiral Four Point Functions}
\vskip  6pt
\vskip 2.0 true cm
\centerline {F.A. Dolan and H. Osborn${}^\dagger$}

\vskip 12pt
\centerline {\ Department of Applied Mathematics and Theoretical Physics,}
\centerline {Wilberforce Road, Cambridge CB3 0WA, England}
\vskip 1.5 true cm

{\eightpoint
\parindent 1.5cm{

{\narrower\smallskip\parindent 0pt

The conformal partial wave analysis of four point functions of
$\half$-BPS operators belonging to the $SU(4)$ $[0,p,0]$ representation
is undertaken for $p=2,3,4$. Using the results of $\N=4$ superconformal
Ward identities the contributions from protected short and semi-short 
multiplets are identified in terms of the free field theory. In the large $N$
limit contributions corresponding to long multiplets with twist up to $2p-2$
are absent. The anomalous dimensions for twist two singlet multiplets are 
found to order $g^4$ and agree with other perturbative calculations. Results
for twist four and six are also found.



\narrower}}

\vfill
\line{${}^\dagger$ 
address for correspondence: Trinity College, Cambridge, CB2 1TQ, England\hfill}
\line{\hskip0.2cm emails:
{{\tt fad20@damtp.cam.ac.uk} and \tt ho@damtp.cam.ac.uk}\hfill}
}

\eject}
\pageno=1

\newsec{Introduction}

The spectrum of operators and their couplings in a conformal field theory
can be explored by analysing the four point correlation functions
for any basic set of operators $\phi_i$ such that 
$\langle \phi_1 \phi_2 \phi_3 \phi_4 \rangle$ can be calculated.
Using the operator product expansion for any pair $\phi_i \phi_j$
which appear in the correlation function provides an expansion in terms
of conformal partial waves, functions which depend on the spins and scale
dimensions of the operators which are present in the operator product expansion. 
Just as many of the particles appearing in the data tables were found by a 
partial wave analysis of experimentally measured scattering amplitudes then 
the spectrum and anomalous dimensions of operators may be determined through 
the conformal partial wave expansion of conformal four point functions.

For the simplest example we may consider a single scalar field $\phi$ of
scale dimension $\Delta_\phi$ so that
\eqn\pppp{
\l \phi(x_1) \phi(x_2) \phi(x_3) \phi(x_4) \r
= { 1 \over \big ( x_{12}^{\, 2}\, x_{34}^{\, 2} \big )^{\De_\phi}} \,
\G(u,v) \, ,
}
for
\eqn\defuv{
x_{ij} = x_i - x_j \, , \qquad
u = { x_{12}^{\, 2}\, x_{34}^{\, 2} \over x_{13}^{\, 2}\, x_{24}^{\, 2} }
\, ,\quad
v = { x_{14}^{\, 2}\, x_{23}^{\, 2} \over x_{13}^{\, 2}\, x_{24}^{\, 2} }
\, ,
}
where $u,v$ are conformal invariants. The operator product expansion
here takes the form
\eqn\OPEpp{
\phi (x_1) \phi(x_2) = \sum_{\Delta,\ell,I} C_{\phi\phi O^I}\, 
{1\over \big ( x_{12}^{\, 2}\big ) ^{{1\over 2}(2\De_\phi -\Delta+\ell)}}\, 
C^{(\ell)}_\Delta(x_{12}, \pr_{x_2})_{\mu_1 \dots \mu_\ell}
O^I{}_{\!\mu_1 \dots \mu_\ell}(x_2) \, ,
}
where $O^I{}_{\!\mu_1 \dots \mu_\ell}$ is a symmetric traceless rank $\ell$ 
tensor conformal primary operator with scale dimension $\Delta$, $I$ labels 
different 
operators with the same $\Delta,\ell$. $\{O^I{}_{\!\mu_1 \dots \mu_\ell}\}$ 
are assumed to form the complete set of operators
appearing in the operator product expansion of $\phi\phi$. In \OPEpp\
$C^{(\ell)}_\Delta (x, \pr)$ are differential operators constructed so
that \OPEpp\ is compatible with form of the three point function
$\langle \phi \phi O^I \rangle$ and the two point function
$\langle O^I O^J \rangle$. The first is given by 
\eqn\ppO{
\l  \phi (x_1)  \phi(x_2) \, O^I{}_{\!\mu_1 \dots \mu_\ell}(x_3) \r 
= {1\over C_{O^I}}  C_{\phi\phi O^I} \, 
{1\over \big ( x_{12}^{\, 2}\big ) ^{\De_\phi}} \Big (
{ x_{12}^{\, 2} \over x_{13}^{\, 2} \, x_{23}^{\, 2} } \Big )^{{1\over 2}
(\De - \ell)}  Z_{\{\mu_1} \dots Z_{\mu_\ell\}}  \, , 
}
where
\eqn\defZ{
Z_\mu = {x_{13\mu}\over x_{13}^{\, 2} }-{x_{23\mu}\over x_{23}^{\, 2}} \, , 
}
and $\{\dots \}$ denotes symmetrisation and removal of traces. 
The two point function is also 
\eqn\OO{
\l O^I{}_{\!\mu_1 \dots \mu_\ell} (x_1) \, O^J{}_{\!\nu_1 \dots \nu_\ell}(x_2) 
\r = C_{O^I} \, \de^{IJ} {1\over \big ( x_{12}^{\, 2}\big ) ^\De} \, 
I_{\{ \mu_1 | \nu_1}(x_{12}) \dots I_{\mu_\ell \} \nu_\ell}(x_{12})\, , 
}
where $I_{\mu\nu}$ is the inversion tensor,
\eqn\Inv{
I_{\mu\nu}(x) = \de_{\mu\nu} - 2 {x_\mu x_\nu \over x^2} \, .
}
The differential operator $C^{(\ell)}_\Delta (x, \pr)$ is then such that
$C^{(\ell)}_\Delta (x, 0)_{\mu_1 \dots \mu_\ell} = x_{\{\mu_1} \dots
x_{\mu_\ell \}}$. 

Using the operator product expansion \OPEpp\ in \pppp\ gives rise to
the conformal partial wave expansion,
\eqn\opeG{
\G(u,v) = \sum_{\Delta,\ell} a_{\ell}^\Delta \,
u^{{1\over 2}(\Delta-\ell)} G_{\Delta}^{(\ell)}(u,v) \, ,
}
where
\eqn\apart{
a_{\ell}^\Delta = \sum_I {1\over C_{O^I}} \big ( C_{\phi\phi O^I} \big )^2 \, ,
}
and the partial wave amplitudes $G_{\Delta}^{(\ell)}(u,v)$ are explicitly 
known functions, at least in dimensions $d=2,4,6$, which may be expanded in 
powers of $u$ and $1-v$, satisfying,
\eqn\Gsym{
G_{\Delta}^{(\ell)}(u,v) = (-1)^\ell v^{-{1\over 2}(\Delta-\ell)}
G_{\Delta}^{(\ell)}(u/v,1/v) \, .
}
For functions $\G$ satisfying crossing symmetry under $x_1 \leftrightarrow x_2$,
$\G(u,v)= \G(u/v,1/v)$, this ensures that only
$\ell$ even appears in the summation in \opeG. 
In general we may set $C_{O^I}=1$ by a choice of normalisation. 
We also assume $C_\phi=1$. 
For the energy momentum tensor, for which $\Delta=d, \ell=2$, there is
however a canonical normalisation such that \refs{\Cardy,\hughone}
\eqn\CT{
C_{\phi\phi T} = - {1\over S_d} \, {\De_\phi d\over d-1} \, , \qquad
S_d = {2 \pi^{{1\over 2}d} \over \Gamma(\half d)} \, .
}

In conformal field theories the essential function $\G(u,v)$ in \pppp\
may be calculated as a perturbative series in a small parameter
$\eps$, either the coupling or $1/N$ for large $N$, and
in general it has an expansion in $\eps$ of the form
\eqn\expG{
\G(u,v) = \G_0(u,v) + \sum_{r=1,2,\dots } \!\! \eps^r \, \sum_{s=0}^r \ln^s \! u\
\G_{r,s}(u,v) \, ,
}
where $\G_0$ is the free field contribution. The conformal partial
wave expansion for $\G$ may be written as in \OPEpp\ with $\De \to \De_I$
where $I$ now labels different operators with the same scale dimension
$\De_0$ and $\ell$ for $\eps=0$. With the expansions
\eqn\expD{
\De_I = \De_0 + \eps \, \De_{I,1}  + \eps^2 \De_{I,2} + \dots \, , 
\quad a_\ell^{\De_I} = a_{\ell,I}^{\De_0} + \eps \, b_{\ell,I}^{\De_0}
+ \dots \, ,
}
we may derive 
\eqn\importtwo{\eqalign{
\G_{0}(u,v) = {}& \sum_{\De_0,\ell,I} a^{\De_0}_{\ell,I} \,
u^{{1\over 2}(\De_0-\ell)}G^{(\ell)}_{\De_0} (u,v) \, , \cr
\G_{1,1}(u,v) = {}& \half \sum_{\De_0,\ell,I } \De_{I,1} \,
a^{\De_0}_{\ell,I} \, 
u^{{1\over  2}(\De_0-\ell)}G^{(\ell)}_{\De_0}(u,v) \,,\cr
\G_{1,0}(u,v) = {}& \sum_{\De_0,\ell,I} u^{{1\over 2}(\De_0-\ell)}
\Big( b^{\De_0}_{\ell,I} \,G^{(\ell)}_{\De_0} (u,v) 
+\De_{I,1}\,a^{\De_0}_{\ell,I} \, 
G^{(\ell)}_{\De_0}{}' (u,v)  \Big)\,,\cr
\G_{2,2}(u,v) = {}& {\textstyle{1\over 8}}\sum_{\De_0,\ell,I}
\De_{I,1}^2\,a^{\De_0}_{\ell,I}\, 
u^{{1\over 2}(\De_0-\ell)}  G^{(\ell)}_{\De_0}(u,v) \,,\cr
\G_{2,1}(u,v) = {}& \half \sum_{\De_0,\ell,I}
u^{{1\over 2}(\De_0-\ell)}
\Big( \big (\De_{I,2}  \, a^{\De_0}_{\ell,I} + \De_{I,1} \, 
b^{\De_0}_{\ell,I} \big ) G^{(\ell)}_{\De_0}(u,v)  +
\De_{I,1}^2\,a^{\De_0}_{\ell,I} \, G^{(\ell)}_{\De_0}{}'(u,v) \Big)\,,\cr}
}
where
\eqn\Gde{
G^{(\ell)}_{\De_0}{}'(u,v) = {\pr \over \pr \De} G^{(\ell)}_{\De}(u,v)
\Big | _{\De= \De_0} \, .
}
In principle, for expansions to arbitrary orders in $\eps$, we may determine
$\sum_I \De_{I,r}^{n_{\vphantom g}} \, a^{\De_0}_{\ell,I}$. For $r=1$ for
instance we need to consider the expansion of the coefficient of
$( \eps \, \ln u )^n $, $\G_{n,n}(u,v)$. However 
this is an inefficient procedure even for determining $\De_{I,1}$ except if 
there is no degeneracy and it is sufficient to restrict to $n=1$.
Such perturbative expansions were first applied for conformal $O(N)$
sigma models in dimensions $2<d<4$ \refs{\Lang,\Pet} but were later applied 
to superconformal gauge theories in four dimensions 
\refs{\Hokone,\OPEN,\OPEW,\Edent,\Except,\Bia,\HMR}.

In this paper we apply such expansions to the four point correlation
functions of $\half$-BPS operators, whose lowest scale dimension
operator belongs to the $SU(4)$ representation with Dynkin labels $[0,p,0]$,
in $\N=4$ superconformal theories, extending previous results and offering
a more complete discussion.
For such correlation functions the implications of $\N=4$ superconformal 
symmetry have been analysed \refs{\Howe,\ASok,\NO}
and it has been shown how this is compatible with the appearance of various 
possible supermultiplets in the operator product expansion \NO. In addition
for $p=2,3,4$ perturbative results to order $g^4$ in the gauge coupling
have been found \refs{\pert,\Kon,\hpert} and also to order $1/N^2$ using the 
AdS/CFT correspondence \refs{\Arut,\ADHS,\Degen}.
We use these results in our discussion. A special feature of
superconformal theories is the existence of supermultiplets which  satisfy
shortening conditions and in consequence the operators have no anomalous
scale dimensions. Only long multiplets, where the dimension of conformal
primary operators is proportional to $2^{16}$ in $\N=4$, can have 
anomalous dimensions but in order to determine these it is necessary
to separate off the contributions of short, for which the lowest scale
dimension or superconformal primary operator has $\ell=0$, and
semi-short multiplets, for which $\ell >0$ is allowed. This may be accomplished 
using the solutions of the superconformal Ward identities.

In detail in section 2 we show how the results of superconformal symmetry,
following \NO, allow an operator product expansion for the correlation functions
of $\N=4$ chiral primary operators in which the contribution of 
entire supermultiplets is evident. It is shown how non unitary multiplets 
are cancelled and how there are potential ambiguities due to the decomposition
of a long multiplet into semi-short multiplets at unitarity threshold. In section
3 this is applied to free field theory, which is initially expressed in terms
of a general linear combination of symmetric polynomials and later specialised
to the large $N$ limit. The operator product expansion is carried out in detail
and contributions for various supermultiplets identified. In section 4
results from the AdS/CFT correspondence are applied in the large $N$ strong
coupling limit. It is shown how low twist long multiplets which appear in the
conformal wave expansion of the large $N$ results without anomalous dimensions
cancel exactly corresponding contributions from free field theory. This is in 
accord with the expectation from string theory that these should  decouple \Gub. 
In section 5 some perturbative results are considered. Some technical details 
relevant for obtaining the operator product expansions are left to five appendices.

\newsec{Superconformal Expansions}

In a $\N=4$ superconformal theory the correlation functions for 
chiral primary $\half$-BPS operators, 
which belong to the $SU(4)$ $[0,p,0]$ representation,  are given by 
symmetric traceless tensor fields $\vphi_{r_1 \dots r_p}(x)$ and 
have $\Delta =p$ and $\ell=0$. Their correlation functions may be calculated 
both perturbatively and for large $N$ through the AdS/CFT correspondence. 
For detailed analysis for arbitrary $p$ it is very convenient to consider
instead $\vphi^{(p)}(x,t) = \vphi_{r_1\dots r_p}(x)\, t_{r_1} \dots 
t_{r_p}$
for $t_r$ an arbitrary six dimensional complex null vector. The four point
correlation functions of interest here are then simply given by an extension
of \pppp
\eqn\fourp{
\l \vphi^{(p)} (x_1,t_1) \, \vphi^{(p)}(x_2,t_2)\,
\vphi^{(p)}(x_3,t_3)\, \vphi^{(p)}(x_4,t_4 )\r 
=  \bigg ( {t_1 {\cdot t_2} \, t_3 {\cdot t_4} \over 
x_{12}^{\, 2}\, x_{34}^{\, 2} } \bigg )^{\! p} \G(u,v;\sigma,\tau) \, ,
}
for $u,v$ as in \defuv\ and $\sigma,\tau$ $SU(4)$ invariants which are defined 
by
\eqn\defst{
\sigma = {t_1{\cdot t_3} \, t_2{\cdot t_4} \over
t_1{\cdot t_2} \, t_3{\cdot t_4}} \, , \qquad
\tau  = {t_1{\cdot t_4} \, t_2{\cdot t_3} \over
t_1{\cdot t_2} \, t_3{\cdot t_4}} \, .
}
Necessarily, since the correlation function is homogeneous of degree $p$ 
in each
$t_i$,  $\G(u,v;\sigma,\tau)$ is a polynomial of degree $p$ in $\sigma, 
\,\tau$
(i.e it may be expanded in monomials $\sigma^r \tau^s$ with $r+s\le p$).
It may be decomposed into contributions for the differing $SU(4)$ 
representations
in the tensor product $[0,p,0] \otimes [0,p,0]$ for 
$\vphi^{(p)} (x_1,t_1) \, \vphi^{(p)}(x_2,t_2)$ by writing
\eqn\exG{
\G(u,v;\sigma,\tau) = \sum_{0\le m \le n \le p} \!\! a_{nm} ( u,v)  \, 
Y_{nm}(\sigma, \tau)\, ,
}
where $a_{nm}$ corresponds to the representation $[n-m,2m,n-m]$ and 
$Y_{nm}$ are two variable harmonic polynomials which, as shown in \NO, are 
given explicitly in terms of single variable Legendre polynomials by
\eqn\poly{
Y_{nm}(\sigma,\tau) =
{P_{n+1}(y) P_{m}(\bar y) - P_{m} (y) P_{n+1} (\bar y)
\over y - \bar y} \, , \quad \sigma = \quar ( 1+y) ( 1+ \bar y) \, , \
\tau = \quar ( 1-y) ( 1- \bar y) \, .
}
In a conformal theory the operator product expansion is reflected by the 
expansion of $a_{nm}$ in terms of conformal partial waves which takes the 
form
\eqn\opG{
a_{nm}(u,v) = \sum_{\Delta,\ell} a_{nm,\ell}^\Delta \, 
u^{{1\over 2}(\Delta-\ell)} G_{\Delta}^{(\ell)}(u,v) \, ,
}
where $a_{nm,\ell}^\Delta$ corresponds to the contribution of a conformal
primary operator with spin $\ell$ and scale dimension $\Delta$. The 
conformal partial wave functions $ G_{\Delta}^{(\ell)}(u,v)$ are explicitly 
known functions which have a simple expression in four dimensions. 
Of course as a consequence of 
superconformal symmetry the conformal primary operators must belong to 
supermultiplets in each of which there is a finite range of related 
possible $\ell$ and $\Delta$, the operator with lowest $\Delta$ is termed
the superconformal primary for the appropriate supermultiplet. Manifestly 
all operators belonging to a given supermultiplet must have the same anomalous 
dimension. In general it is non trivial to separate the contributions 
in the operator product expansion of descendant operators from superconformal 
primary operators.

This difficulty is easily resolved  by considering the solution of the 
superconformal Ward identities for $\G$ which require \refs{\ASok,\NO}
\eqn\Ward{
\G\big (x\bar x,(1-x)(1-\bar x); \alpha \bar \alpha , 
(1-\alpha)(1- \bar \alpha ) \big ) \big |_{\bar \alpha = {1\over \bar x}} 
= k
+ \Big ( \alpha - {1\over x} \Big ) \hat f ( x, 2\alpha -1 ) 
\,  .
}
The solution of \Ward\  can be written as
\eqn\solW{\eqalign{
\G(u,v;\sigma,\tau) ={}& k + \G_{\hat f}(u,v;\sigma,\tau) + 
s(u,v;\sigma,\tau)
\H(u,v;\sigma,\tau) \, , \cr
s(u,v;\sigma,\tau) = {}& v + \sigma^2 uv + \tau^2 u + \sigma \, v(v-1-u) + 
\tau (1-u-v) + \sigma \tau \, u(u-1-v) \, , \cr}
}
where $\G_{\hat f}$ may be explicitly given in terms of $\hat f$. The 
constant $k$ and the function $\hf(x,y)$, a polynomial in $y$ of 
degree $p-1$, are determined by the free field results for $\G$
whereas dynamical effects, which lead to anomalous dimensions, are 
contained in the function $\H$ which is a polynomial in $\sigma,\tau$ 
of degree $p-2$. Instead of considering the conformal partial wave expansion
of $\G$ it is sufficient to expand $\H$ and $\hf$ as in \NO\ so that
\eqn\exH{\eqalign{
\H(u,v;\sigma,\tau) = {}& \sum_{0\le m \le n\le p-2} A_{nm}(u,v) 
\, Y_{nm}(\sigma,\tau) \cr
= {}& \sum_{{0\le m \le n\le p-2\atop \Delta,\ell}} 
A^\Delta_{nm,\ell} \, u^{{1\over 2}(\Delta-\ell)}
G_{\Delta +4}^{(\ell)}(u,v) \, Y_{nm}(\sigma,\tau) \, . \cr}
}
As a consequence of crossing symmetry of \fourp\ under $x_1,t_1
\leftrightarrow x_2,t_2$ we have
\eqn\AA{
A_{nm}(u,v) = (-1)^{n+m} {1\over v^2} A_{nm}(u/v,1/v) \, ,
}
and hence, from \Gsym, in the conformal partial wave expansion \exH\
we have $\ell=0,2,\dots $ for  $n+m$ even and otherwise $\ell=1,3, \dots$. 
We may also similarly expand $\hf$ in the form
\eqn\exf{
\hf(x,y) = -2\!\! \sum_{{0\le n\le p-1\atop\ell}} b_{n,\ell} \, 
g_{0,\ell+2}(x)\,  P_n(y) \quad 
\cases{&$\ell$ odd if $n$ even\cr&$\ell$ even if $n$ odd} \, , 
}
with the definition
\eqn\gell{
g_{t,\ell}(x) = 
\big ({-\half }x\big )^{\ell} F\big (t+\ell,t+\ell;2t+2\ell;x\big ) \, . 
}
In general in \exf\ $\ell \ge 0$ except when $n=0$ it is necessary in 
general 
to include $\ell=-1$ in the sum. These expansions determine the conformal
partial wave expansion for $a_{nm}(u,v)$ in a form where the contribution
of each superconformal multiplet is explicit,
\eqn\anm{\eqalign{
a_{n'm'} = {}& k \, \delta_{n'0}  \delta_{m'0} +
\sum_{{0\le m \le n\le p-2\atop \Delta,\ell}} \!\!\!
A^\Delta_{nm,\ell}\, a_{n'm'}\big (\A^\Delta_{nm,\ell}\big ) \cr
&{} + {1\over 4} \sum_\ell b_{0,\ell+1} \, a_{n'm'}\big (\C_{00,\ell}\big 
) 
+ \sum_{{0\le n\le p-2\atop\ell}} \!\! b_{n+1,\ell} \, a_{n'm'}
\big (\D_{n0,\ell}\big ) \, . \cr}
}
Here $a_{nm}(\M)$ are the contributions corresponding to a supermultiplet
$\M$, for each $\M$ $a_{nm}(\M)$ is given by a finite linear combination
of $u^{{1\over 2}(\Delta-\ell)} G_{\Delta}^{(\ell)}(u,v)$ for the 
appropriate $\Delta,\ell$ corresponding to all operators belonging to 
the $SU(4)$ representation labelled by $nm$ in the multiplet $\M$. 
For each possible $\N=4$ supermultiplet, as described for instance in \Short,
detailed results for $a_{nm}(\M)$ are given in \NO.
$\A^\Delta_{nm,\ell}$ denotes a generic long multiplet whose
lowest state has scale dimension $\Delta$, spin $\ell$ and belongs to
the $SU(4)$ representation labelled by $nm$, for unitarity $\Delta \ge 
2n+\ell+2$.
Conversely $\D_{nm,\ell}$ and $\C_{nm,\ell}$ are semi-short 
supermultiplets in which 
the scale dimension for the lowest state is determined to be $2m+\ell$ and 
$2n+\ell+2$ respectively. These occur in the decompositions
\eqn\multD{\eqalign{
\A^{2m+\ell}_{nm,\ell} \simeq {}& \D_{nm,\ell} \oplus \D_{n\, 
m{+1},\ell{-1}}
\oplus \dots  \, , \qquad m=0,1,\dots, n-1 \, , \cr
\A^{2n+\ell}_{nn,\ell} \simeq {}& \D_{nn,\ell} \oplus 
\C_{n{+1}\, n{+1},\ell{-2}} \oplus \dots \, , \cr
\A^{2n+\ell}_{nm,\ell} \simeq {}& \C_{nm,\ell} \oplus 
\C_{n{+1}\, m,\ell{-1}} \oplus \dots  \, , \qquad\quad
n=m,m+1,\dots  \,  \, , \cr}
}
where in each case two multiplets are omitted which are irrelevant since
they cannot contribute in the operator product expansions here.
Correspondingly we have\foot{For convenience the normalisation of
$ a_{n'm'}^{\vphantom g}( \A^\Delta_{n m,\ell}  )$ is changed from
that used in \NO\ by a factor 16.}
\eqna\CD
$$\eqalignno{
a_{n'm'}^{\vphantom g}\big ( \A^{2m+\ell}_{n m,\ell} \big )
= {}&  a_{n'm'}^{\vphantom g} \big ( \D_{n m,\ell} \big ){}
+ {m+1\over 4(2m+1)} \, a_{n'm'}^{\vphantom g}
\big ( \D_{n\,m{+1},\ell-1} \big ) \, , & \CD{a} \cr
a_{n'm'}^{\vphantom g}\big ( \A^{2n+\ell}_{n n,\ell} \big )
= {}&  a_{n'm'}^{\vphantom g} \big ( \D_{n n,\ell} \big ){}
+ {(n+1)(n+2)\over 16 (2n+1)(2n+3)} \,
a_{n'm'}^{\vphantom g} \big ( \C_{n+1\,n+1,\ell-2} \big ) \, , & \CD{b} 
\cr
a_{n'm'}^{\vphantom g}\big ( \A^{2n+\ell+2}_{n m,\ell} \big )
= {}&  a_{n'm'}^{\vphantom g} \big ( \C_{n m,\ell} \big ){}
+ {n+2\over 4(2n+3)} \, a_{n'm'}^{\vphantom g}
\big ( \C_{n{+1}\,m,\ell-1} \big ) \, .  & \CD{c} \cr}
$$
When $\ell=0$ in \CD{b}\  we may use
\eqn\CB{\eqalign{
a_{n'm'}\big (\C_{n n,{-2}}\big ) = {}& - 4 \, a_{n'm'}\big (\B_{nn}\big ) 
+
{(n+1)(n+2)\over (2n+1)(2n+3)}\, a_{n'm'}\big (\B_{n{+1}\,n{+1}}\big ) \, 
, \cr
a_{n'm'}\big (\B_{00}\big ) = {}& \delta_{n'0}  \delta_{m'0} \, ,\cr}
}
where, for $n\ge 1$, $\B_{nn}$ denotes the $\half$-BPS short multiplet
corresponding to the $[0,2n,0]$ $SU(4)$ representation with $\Delta=2n$ 
($\B_{00}$ is the trivial singlet multiplet for the identity operator).
In \CD{c}\ for $\ell=0$ we also have
\eqn\aCB{
a_{n'm'}^{\vphantom g}\big ( \C_{n m,-1} \big ) = {n+1\over 2n+1} \,
a_{n'm'}^{\vphantom g}\big ( \B_{n+1\, m} \big ) \, ,
}
where, for $n>m$, $\B_{nm}$ is a $\quar$-BPS short multiplet corresponding 
to the  $[n-m,2m,n-m]$ representation.

The multiplets $\D_{n m,\ell}$ are non unitary and their contributions
must be cancelled in the conformal partial wave expansion for a unitary 
theory. The coefficients for non unitary long multiplets with $\Delta =
2t+\ell$, $\ta = 0,1,\dots, n$, are written
$A^{2t+\ell}_{nm,\ell} \equiv  A_{nm,t\ell}$ and we must then have
\eqn\bA{\eqalign{
b_{n+1,\ell}+A_{n0,0\ell} = {}& 0 \, , \ \ n=0,1,\dots, p-2 \, , \cr
{m+1\over 4(2m+1)} A_{nm,m\ell} + A_{n\, {m+1},m+1\,\ell-1} = {}& 0 \, , \ 
\
m=0,1,\dots , n-1 \le p-3 \, . \cr}
}
In other cases we must also require
\eqn\UA{
A_{nm,t\ell} = 0 \, , \qquad t = 0, \dots , n \, , \ \ t \ne m \, .
}
With the above relations  we get
\eqn\exaa{\eqalign{
a_{n'm'} = {}& k \, \delta_{n'0}  \delta_{m'0} +
\sum_{{0\le m \le n\le p-2\atop \Delta,\ell}} \!\!\!
A^\Delta_{nm,\ell} \, a_{n'm'}\big (\A^\Delta_{nm,\ell}\big )
+ {1\over 4}  \sum_\ell  b_{0,\ell+1} \, a_{n'm'}\big (\C_{00,\ell}\big )  
\cr
&{}  + \sum_{{0\le n\le p-2\atop\ell}} \!\!  {(n+1)(n+2)\over 
16(2n+1)(2n+3)}\,
A_{nn,n\ell} \, a_{n'm'}\big (\C_{n{+1}\,n{+1},\ell{-2}}\big ) \, . \cr}
}
As may be seen from \CD{c}\ the contributions to the partial waves of the 
semi-short multiplets $\C_{nn,\ell}$ for $n=0,1,\dots, p-1$ cannot be 
combined to those corresponding to a long multiplet. However
using \CD{c}\ any particular semi-short multiplet can be taken to be part
of a long multiplet at the expense of introducing the contribution for
another semi-short multiplet. In an interacting theory in general all
long multiplets are expected to gain anomalous dimensions but in a free 
theory with canonical scale dimensions there is an inherent ambiguity as a 
consequence of \CD{c}\ in writing the superconformal partial wave expansion. 
Necessarily, so long as the coefficients are non zero, the contributions 
corresponding to at least $p$ semi-short multiplets for each appropriate 
$\ell$  must be present in  the expansion.
In any interacting theory no contribution corresponding to $\C_{00,\ell}$ 
should be present as it contains higher spin conserved currents. If this is 
removed by 
using \CD{c}\ to be part of the associated twist 2 long multiplet 
$\A^{\ell +2}_{00,\ell}$ we obtain from \exaa
\eqn\exab{\eqalign{
a_{n'm'} = {}& C \, \delta_{n'0}  \delta_{m'0} + 
\sum_{n=1}^p C_n \, a_{n'm'}\big (\B_{nn}\big ) + 
C_{20} \, a_{n'm'}\big (\B_{20}\big ) \cr
&{} + \! \sum_{{0\le m \le n\le p-2\atop \Delta,\ell}} \!\!\!
{\hat A}^\Delta_{nm,\ell} \, a_{n'm'}\big (\A^\Delta_{nm,\ell}\big ) 
- {1\over 24}  \sum_{\ell = 1,3,\dots }  b_{0,\ell+2} \
a_{n'm'}\big (\C_{10,\ell}\big ) \cr 
&{} +  \! \sum_{{0\le n\le p-2\atop\ell = 0,2,\dots}}\!\!
{(n+1)(n+2)\over 16(2n+1)(2n+3)}\,
A_{nn,n\,\ell{+2} } \ a_{n'm'}\big (\C_{n{+1}\,n{+1},\ell}\big ) \, , \cr}
}
where we define for $\Delta = \ell +2$,
\eqn\Aha{
{\hat A}_{00,1\ell} = A_{00,1\ell} + \quar \, b_{0,\ell+1} \, ,
\qquad \ell = 0,2,\dots  \, ,
}
with otherwise ${\hat A}_{nm,t\ell} = A_{nm,t\ell}$.
The remaining coefficients are given by
\eqn\Ck{
C = k -  b_{0,-1}  \, ,  \qquad C_1 = {\ts {1\over 6}} 
( b_{0,-1} - A_{00,00} ) \, ,
}
and for $n=2,3,\dots, p$
\eqn\Cn{\eqalign{
C_n =&{}  - {n(n+1)\over 4(2n-1)(2n+1)} \, A_{n{-1}\, n{-1},n{-1}\,0}\cr
&{} + {(n-1)n^2(n+1)\over 16(2n-3)(2n-1)^2(2n+1)} \, 
A_{n{-2}\, n{-2},n{-2}\,0} \, , \cr}
}
and also, 
\eqn\Cs{
C_{20} = -{\ts {1\over 36}} b_{0,1} \, .
}
For unitarity it is necessary that all coefficients are positive and 
that only contributions for $\Delta \ge 2n + \ell + 2$ arise. 
In general further rearrangements
are necessary to achieve this as demonstrated later. 

\newsec{Free Field Results}

We here consider the superconformal expansions for four point functions 
of BPS operators for $p=2,3,4$ in the free field case. We present the
results in each case first and then draw more general conclusions.

The results for $\G$ in free field theory are expressible in terms of 
crossing
symmetric polynomials $S_p(\sigma,\tau)$ which are defined by
\eqn\Sp{
S_p(\sigma,\tau) = S_p(\tau,\sigma) = \tau^p S_p(\sigma/\tau,1/\tau) \, .
}
A basis for minimal polynomials is given for $p=1,2,3,\dots$ by first
defining
\eqn\Sbasis{
S_{p,i}(\sigma,\tau) = \cases{\sigma^p + \tau^p + 1 \, , &$i=0\,$,\cr
\sigma^{p-i}\tau^i+ \sigma^i \tau^{p-i} + \sigma^{p-i} + \tau^{p-i}
+ \sigma^i + \tau^i \, , &$i=1,2,\dots, \ i<  \half p\,$, \cr
\sigma^{{1\over 2}p}\tau^{{1\over 2}p}+ \sigma^{{1\over 2}p}+
\tau^{{1\over 2}p} \, , &$i=\half p$, $p$ even$\,$. \cr}
}
Assuming
\eqn\Sz{
S_{0,0}(\sigma,\tau) = 1 \, ,
}
we then define a complete set of crossing symmetric polynomials by
\eqn\Sgen{
S_{p(i,j)}(\sigma,\tau) = (\sigma \tau)^j S_{p-3j,i}(\sigma,\tau) \, ,
\qquad i=0,1, \dots , [\half (p-3j)] , \ j =0,1, \dots, [{\ts {1\over 
3}}p]\, ,
}
so that $2i+3j \le p$. For each $p$ a formula for the number of independent 
$(i,j)$ is given in \refs{\NO,\Howe}.
The contributions represented by each polynomial correspond to the different possible
sets of crossing symmetric free field graphs where the vertices are linked by
$l,m,n$ lines, as shown in Fig. 1, where $l+m+n=p$ and $l \ge m \ge n \ge 0$
and where we identify $n=j, \, m = i+j$.

{\midinsert
\hfil \epsfxsize=0.5\hsize
\epsfbox{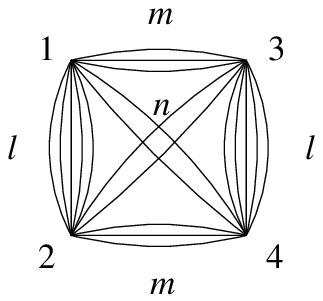} \hfil
\vskip -6pt
{\eightpoint{
\parindent 1cm

{\narrower\smallskip\noindent
Fig. 1 Free field contributions to four point function for $l=5, \, m =3, \, n=2$. 
\smallskip}

\narrower}}
\endinsert}

For $p=2$ there are just two possible polynomials and the  free field results are 
in general expressible as
\eqn\Stwo{
\G_0(u,v;\sigma,\tau) = 
S_{2(0,0)}(\sigma u,\tau u/v) + a \, S_{2(1,0)}(\sigma u,\tau u/v) \, ,
}
for some coefficient $a$. The first term corresponds to disconnected 
graphs, its
coefficient is one as a consequence of normalising the two point function
for the BPS operators to one. In the large $N$ limit the connected 
contribution
is suppressed and we have
\eqn\atwo{
a = {4\over N^2} \, .
}
{}From this we may determine
\eqn\kHtwo{
k = 3 ( 1 + a  ) \, , \quad 
\H_0(u,v;\sigma,\ta) = 1 + {1\over v^2} + a  \, {1\over v} \, ,
}
and
\eqn\ftwo{
\hf(x,y) = \half y \big ( x^2 + x'{}^2  - a  (x+ x') \big ) + 
\half \big ( x^2 - x'{}^2  + ( 2 + 3 a ) (x - x') \big )  \, ,
}
where
\eqn\xx{
x' = {x\over x-1} \, , \qquad x+x' = xx' = {x^2 \over x-1} \, .
}
The expansion of $\H_0$ in \kHtwo\ gives
\eqn\Atwo{
A_{00,t\ell}  =  2^{\ell+1}{(\ell+\ta+1)!^2(\ta!)^2
\over (2\ell+2\ta+2)!(2\ta)!}\,\big ( (\ell+1)(\ell+2\ta+2) + 
a (-1)^\ta \big ) \, , 
}
and for $\hf$ in \ftwo
\eqn\btwo{\eqalign{
b_{1,\ell} = {}& -2^{\ell+1}{(\ell+1)!^2 \over (2\ell+2)!}
\big ((\ell+1)(\ell+2)  + a \big ) \, , \quad \ell=0,2 \dots \, ,\cr
b_{0,\ell} = {}& -2^{\ell+1}{(\ell+1)!^2 \over (2\ell+2)!}
\big (\ell(\ell+3)  - 3a \big ) \, , 
\quad\qquad \ell=-1,1,\dots \, .\cr}
}
As required by \bA\ to remove twist zero $A_{00,0,\ell} + b_{1,\ell} = 0$.
The results \Ck, \Cn\ and \Cs\ give 
\eqn\Ctwo{
C=1 \, , \qquad  C_1 = {\ts {1\over 3}}a  \, , \qquad 
C_2 = {\ts {1\over 30}} \big ( 1 + \half a  \big ) \, , \qquad
C_{20} = {\ts{2\over 27}} \big ( 1 - {\ts {3\over 4}} a \big ) \, .
}
The result that $C=1$ is necessary for consistency with our normalisation
since the coefficient of the identity in the operator product expansion
is the same as that for the two point function. From \Aha\ we get
\eqn\Afztwo{
{\hat A}_{00,1\ell}
= 2^{\ell+1} {(\ell+2)!^2\over (2\ell+4)!} \, a \, .
}
The expansion for the free theory then becomes
\eqnn\extwo
$$\eqalignno{ \!\!\!
a_{n'm'} = {}& \delta_{n'0}  \delta_{m'0} + 
\sum_{n=1}^2 C_n \, a_{n'm'}\big (\B_{nn}\big ) + C_{20}\,a_{n'm'}\big 
(\B_{20}\big )
+ \sum_{\ell=0,2,\dots , \ta\ge 1} 
{\hat A}_{00,t\ell} \, a_{n'm'}\big (\A^{2\ta+\ell}_{00,\ell}\big ) \cr
& {} + {\ts {1\over 24}} \sum_{\ell=0,2,\dots} \Big (
A_{00,0\,\ell{+2} } \, a_{n'm'}\big (\C_{11,\ell}\big )
- b_{0,\ell{+3} } \, a_{n'm'}\big (\C_{10,\ell{+1}}\big ) \Big ) \, . & 
\extwo \cr}
$$
The positivity requirements are satisfied for $0 < a \le {4\over 3}$ (in 
general
for $\N=4$ $SU(N)$ superconformal theory $a=4/(N^2-1)$ so that 
$a={4\over 3}$ when $N=2$).

For $p=3$ the free field theory result for the correlation function
is in general
\eqn\Sthree{
\G_0(u,v;\sigma,\tau) = 
S_{3(0,0)}(\sigma u,\tau u/v) + a \, S_{3(1,0)}(\sigma u,\tau u/v) 
+ b \, S_{3(0,1)}(\sigma u,\tau u/v)  \, ,
}
where as before the leading coefficient is taken to be $1$ and for large 
$N$ 
we have
\eqn\athreeo{
a= {9 \over N^2} \, , \qquad b = {18 \over N^2} \, .
}
{}From \Sthree\ we may determine
\eqn\kthree{
k = 3 + 6 a + b \, ,
}
and
\eqnn\threefree
$$\eqalignno{
\H_0(u,v;\sigma,\tau)= & {1\over v^3} \Big ( \half(\sigma+\tau) u (1+v^3)
+ \half(\sigma- \tau) \big ( -3u(1-v^3) + 2(1-v)(1+v^3) \big ) \cr
&\qquad {}+ u (1+v^3) - 1 + 2v+2v^3 - v^4 \Big )  \cr
&{}+ a \, {1\over v^2} \Big ( \half(\sigma+\tau) u(1+v)
+ \half(\sigma- \tau) (1-v)\big (2(1+v)-u\big ) + (1+v)^2 \Big )  \cr
&{}+ b \, {1\over v} \, , & \threefree \cr}
$$
and
\eqnn\fthree
$$\eqalignno{
\!\! \hf(x,y) = {}& \quar \big ( (y^2+1) (x^3 - x'{}^3) + 2y  
( x^3 + x'{}^3 + x^2 + x'{}^2  )  
+ 2  ( x^2 - x'{}^2 )  + 4(x-x') \big ) \cr
&{} - \quar a \, \big ( (y^2-3) (x^2 - x'{}^2) - 2y (
x^2 + x'{}^2 -2(x+x') ) - 12 (x-x') \big )  \cr
&{}- \half b \, \big ( y(x+x') - (x-x') \big ) \, . & \fthree \cr}
$$
Using  from \poly
\eqn\Yone{
Y_{00} = 1 \, , \qquad Y_{10}= 3(\sigma-\tau) \, , \qquad
Y_{11}= 3(\sigma+\tau) - 1 \, ,
}
and letting
\eqn\Aathree{
A_{nm,t\ell}  = 2^{\ell-2}\,
{(\ell+\ta+1)!^2\, (\ta!)^2 \over 3
(2\ell+2\ta+2)!\, (2\ta)!} \, a_{nm,t\ell} \, ,
}
we have
\eqn\athree{
\eqalign{
a_{11,t\ell} ={} & 
\ta(\ta+1)(\ell+1)(\ell+2\ta+2)(\ell+\ta+1)(\ell+\ta+2)\cr
&{}+ 4a \big ( (1-(-1)^\ta) (\ell+\ta+1)(\ell+\ta+2) - 
(1+(-1)^\ta) \ta(\ta+1) \big ) \, , \cr
a_{10,t\ell}  = {}& (\ta-1)(\ta+2)(\ell+1)(\ell+2\ta+2)
(\ell+\ta)(\ell+\ta+3) \cr
&{} -  4a  (1+(-1)^\ta) (\ell+1)(\ell+2\ta+2) \, , \cr
a_{00,t\ell} = {} & (\ta-2)(\ta+3)(\ell+1)(\ell+2\ta+2)
(\ell+\ta-1)(\ell+\ta+4) \cr
&{}+ 4 a \big ( (1-(-1)^\ta) (\ell+\ta+1)(\ell+\ta+2) -
(1+(-1)^\ta) \ta(\ta+1) \big ) \cr
&{}+ 24 a  (\ell+1)(\ell+2\ta + 2) + 24(2a+b) (-1)^\ta \, .
\cr}
}
In addition from \fthree\ we obtain
\eqn\bthree{\eqalign{
b_{2,\ell} ={}&  {1\over 3}\, 2^{\ell-1} {(\ell+1)!^2 
\over (2\ell+2)!}\, (\ell+1)(\ell+2)\big ( \ell(\ell+3) + 4a \big ) \, , \cr
b_{1,\ell} ={}&  2^{\ell-1} {(\ell+1)!^2
\over (2\ell+2)!}\, \Big ( (\ell+1)(\ell+2)\big ( (\ell-1)(\ell+4) - 4a \big )
- 8a -4b \Big )  \, , \cr
b_{0,\ell} ={}&  {1\over 3}\, 2^{\ell} {(\ell+1)!^2
\over (2\ell+2)!}\,\Big( (\ell-1)(\ell+4) \big ( \ell(\ell+3) - 8a \big )  
- 12a + 6b \Big ) \, .  \cr}
}
As required by \bA\ the twist zero contributions are cancelled as a 
consequence of
${A_{00,0\ell} + b_{1,\ell}} = 0, \ A_{10,0\ell} + b_{2,\ell} = 0$ while 
removal of twist two contributions from the $[0,2,0]$ partial wave follows 
from
$A_{11,1\,\ell} + \quar A_{10,0\,\ell+1} = 0$ which are satisfied by 
\athree\ and 
\bthree. Furthermore $a_{11,0\ell}=a_{10,1\ell}=0$ in accord with \UA.
Using \Ck\ and \Cn\ gives
\eqn\Cthree{
C=1 \, , \qquad  C_1 = {\ts{1\over 3}}a \, , \qquad 
C_2 = {\ts{1\over 60}}(2a+b) \, , \qquad
C_3 = {\ts{1\over 1050}} ( 1 + a ) \, ,
}
where $C=1$ is necessary for consistency as before. From \Aha\ we also 
have
\eqn\Afzthree{
{\hat A}_{00,1\ell} = 2^{\ell+1} {(\ell+2)!^2\over (2\ell+4)!} \, a 
\, .
}
However $A_{00,0\ell} < 0 $ for $\ell \ge 2$ and $b_{0,\ell}>0$ for
$\ell >1$. These negative semi-short contributions may be absorbed into 
the corresponding long multiplets by letting $A\to {\hat A}$ for
\eqn\Aoo{
{\hat A}_{10,2\ell} = A_{10,2\ell} - {\ts {1\over 24}} \, b_{0,\ell{+2}} 
\, ,
\qquad
{\hat A}_{11,2\ell} = A_{11,2\ell} + {\ts {1\over 24}} \, 
A_{00,0\,\ell+2}
}
which gives for this case
\eqn\Atoo{\eqalign{
{\hat A}_{10,2\ell} = {}&
2^{\ell}\, {(\ell+3)!^2\over 3(2\ell+6)!}
\big ( (\ell+1)(\ell+6) a + 2a -b\big )\, ,\cr
{\hat A}_{11,2\ell} = {}&
2^{\ell} {(\ell+3)!^2\over 3(2\ell+6)!}
\big ( (\ell+3)(\ell+4) a + b \big ) \, .
\cr}
}
The remaining twist 4 contribution is given by
\eqn\Atop{
A_{00,2\ell} = 2^{\ell} {(\ell+3)!^2\over 3(2\ell+6)!}
\big ( (\ell+1)(\ell+6)a + b \big ) \, .
}
The expansion for the free theory then becomes
\eqnn\exthree
$$\eqalignno{ \!\!\!
a_{n'm'} = {}& \delta_{n'0}  \delta_{m'0} + 
\sum_{n=1}^3 C_n \, a_{n'm'}\big (\B_{nn}\big ) 
+ C_{20} \, a_{n'm'}\big (\B_{20}\big )  +  C_{31} \, a_{n'm'}\big 
(\B_{31}\big ) \cr
&{} + \sum_{{0\le m \le n\le 1 \atop \ta,\ell}} \!\!\!
{\hat A}_{nm,t\ell} \, a_{n'm'}\big (\A^{2\ta+\ell}_{nm,\ell}\big ) & 
\exthree \cr
\noalign{\vskip -4pt}
+ {} & 
{\ts {1\over 160}} \!\! \sum_{\ell = 0,2,\dots}\!\!\! \Big ( 
4 A_{11,1\,\ell{+2} } \, a_{n'm'}\big (\C_{22,\ell}\big ) 
-  A_{00,0\,\ell{+4} } \, a_{n'm'}\big (\C_{21,\ell{+1}}\big ) 
+ b_{0,\ell{+3} } \, a_{n'm'}\big (\C_{20 ,\ell}\big ) \Big ) \, , \cr}
$$
with in general
\eqn\Cst{
C_{31} = - {\ts {3 \over 800}} \, A_{00,02} \, .
}
Hence from \Cs\ and \Cst\ we obtain
\eqn\Cqq{
C_{20} = {\ts {1\over 54}}(2a-b) \, , \qquad 
C_{31} = {\ts {3\over 2000}}(18 - 4a-b) \, .
}

For $p=4$ the free field theory result for the correlation function
is in general
\eqn\Sfour{\eqalign{
\G_0(u,v;\sigma,\tau) = {}& 
S_{4(0,0)}(\sigma u,\tau u/v) + a \, S_{4(1,0)}(\sigma u,\tau u/v) \cr
&{} + b \, S_{4(2,0)}(\sigma u,\tau u/v)  + 
c \, S_{4(0,1)}(\sigma u,\tau u/v) \, , \cr}
}
where for large $N$ we have
\eqn\afouro{
a= b = {16 \over N^2} \, , \qquad c = {32 \over N^2} \, .
}
In this case from \Sfour\ we have
\eqn\kfour{
k = 3(1+2a+b+c) \, ,
}
and
\eqnn\fourfree
$$\eqalignno{
\H_0 (& u,v;\sigma,\tau) \cr
= {}& {1\over v^4}  \Big ( \sigma\tau \, u^2(1+v^4) \cr
&\ \ {}  + \half (\sigma-\tau)^2
\big ( 2(1-v)^2(1+v^4) - 5u (1+v^5) + 3uv (1+v^3) + 4u^2(1+v^4) \big ) \cr
&\ \ {}+ \half (\sigma^2 - \tau^2) \big ( u(1-v)(1+v^4) - 2u^2 (1-v^4) 
\big ) \cr
&\ \ {} + \half (\sigma+\tau) 
\big ( - (1-v)^2 + u (1+v) \big ) (1+v^4)  \cr
&\ \ {} + \half (\sigma-\tau) 
\big ( (1-v) (- 3(1+v)+7u) (1+v^4) + 8v(1-v)(1+v^3) - 4u^2 (1-v^4) \big ) 
\cr
&\ \ {}+ 1+v^6 - 3v(1-v)(1-v^3) - 2u (1-v)(1-v^4) + u^2(1+v^4) \Big )  \cr
&{}+  {a\over v^3}  \Big ( \sigma\tau \, u^2(1+v^2) \cr
&\ \ {}  + \half (\sigma-\tau)^2
\big ( 2(1+v^2)^2- 4v(1+v^2) - 3u (1+v^3) + uv (1+v) + u^2(1+v^2) \big ) 
\cr
&\ \ {}+ \half (\sigma^2 - \tau^2) 
\big ( u(1-v)(1+v^2) - u^2 (1-v^2) \big ) \cr
&\ \ {} + \half (\sigma+\tau) 
\big ( - (1+v^2)^2 + 2v(1+v^2) +2u (1+v^3) +uv(1+v) \big )\cr
&\ \ {} + \half (\sigma-\tau) 
\big ( 1-v^4 + 2v(1-v^2) -uv(1-v) - 2u (1-v^3) \big ) \cr
&\ \ {} - (1 - v^2)^2 + 2v(1+v^2)  + u (1 + v^3) \Big )  \cr
&{}+  {b\over v^2}  \Big ( \sigma\tau \, u^2 + \half (\sigma-\tau)^2
\big ( 2(1-v)^2- u (1+v) \big ) + \half (\sigma^2 - \tau^2) u(1-v) \cr
&\ \quad {} + \half (\sigma+\tau)
\big ( - (1-v)^2 + u (1+v) \big ) + \half (\sigma-\tau) (1-v) (1+v-u) \cr
&\ \quad {} + 1 + v + v^2 \Big )  \cr
&{}+  {c\over v^2}  \Big ( \half (\sigma+\tau) u (1+v)  
+ \half (\sigma-\tau) (1-v) \big ( 2(1+v) -u \big ) +  3v \Big ) \, . & 
\fourfree\cr}
$$
In addition
\eqnn\ffour
$$\eqalignno{
\!\! \hf(x,y) = {}& {\ts{1\over 8}} \big ( (y^3+3y) (x^4 + x'{}^4) + y^2
( 3( x^4 - x'{}^4 ) + 2 ( x^3 - x'{}^3 ) ) \big )  \cr
&{} + \half y ( x^3 + x'{}^3 + x^2 + x'{}^2  ) \cr
&{} + {\ts{1\over 8}} 
\big (x^4 - x'{}^4 +  2( x^3 - x'{}^3 ) + 4( x^2 - x'{}^2 ) + 8(x - x')  
\big ) \cr
&{} - {\ts{1\over 8}}  a \, \big (
(y^3-5y)( x^3 + x'{}^3 + x^2 + x'{}^2 +  2(x+x'))
-  y^2 (x^3 - x'{}^3 - 3( x^2 - x'{}^2) ) \cr
&{} - {\ts{1\over 8}}  a \, \big (  16 y  (x+x') 
-3(x^3 - x'{}^3) - 7 (x^2 - x'{}^2) - 24 (x-x') \big ) \, ,  \cr
&{} + {\ts{1\over 8}} b \, \big ( (y^3+3y) (x^2 + x'{}^2 + 2 (x+x') ) - 
y^2
(x^2 - x'{}^2 ) \big )  \cr
&{} - {\ts{1\over 8}} b \, \big ( 12 y(x+x') - 5 ( x^2 - x'{}^2 ) 
- 12 (x-x') \big ) \cr
&{} - \quar c \, \big ( (y^2-1) ( x^2 - x'{}^2 ) + 6y (x+x') - 6 (x-x') 
\big ) \, .
& \ffour \cr}
$$
For this case we take from \poly\ as well as \Yone
\eqn\Ytwo{\eqalign{
Y_{20} = {}& 10(\sigma-\tau)^2 - 5 (\sigma+\tau) + 1 \, , \qquad 
Y_{21}= 10(\sigma^2-\tau^2)  - 5 (\sigma - \tau) \, , \cr
Y_{22}= {}& 10(\sigma^2 + \tau^2) + 40 \, \sigma \tau  - 8(\sigma+\tau) + 
1 \, ,
\cr}
}
and writing
\eqn\Aafour{
A_{nm,t\ell}  = 2^{\ell-5}\,
{(\ell+\ta+1)!^2\, (\ta!)^2 \over 45
(2\ell+2\ta+2)!\, (2\ta)!} \, a_{nm,t\ell} \, ,
}
we have the expansion coefficients
\eqnn\afour
$$ \eqalignno{
a_{22,t\ell} ={} & {\ts {1\over 3}}(\ta-1)\ta(\ta+1)(\ta+2)(\ell+1)
(\ell+2\ta+2)(\ell+\ta)(\ell+\ta+1)(\ell+\ta+2)(\ell+\ta+3)\cr
&{}+ 6 \, a \big ( (1+ (-1)^\ta)\,  t (t+1)(\ell + t ) ( \ell+t+3)
((\ell+1)(\ell+2t+2) + 2 ) \cr
\noalign{\vskip-4pt}
& \hskip 1cm {} + (1 -  (-1)^\ta) (t-1)(t+2) (\ell + t +1) ( \ell+t+2)
((\ell+1)(\ell+2t+2) - 2 ) \big )\cr
&{}+ 24 \, b \big ( (1+ (-1)^\ta)\,  t (t+1)(\ell + t ) ( \ell+t+3) \cr
\noalign{\vskip-4pt}
& \hskip 3cm {} - (1 -  (-1)^\ta) (t-1) (t+2)(\ell + t +1 ) ( \ell+t+2) 
\big )\, , \cr
a_{21,t\ell}  = {}& (\ta-2)\ta(\ta+1)(\ta+3)(\ell+1)(\ell+2\ta+2)
(\ell+\ta-1)(\ell+\ta+1)(\ell+\ta+2)(\ell+\ta+4) \cr
&{}- 36 \, a  (1- (-1)^\ta) (\ell + 1) ( \ell+ 2t+ 2)
\big ( (\ell+3)(\ell+2t) + 2(t+1)(t-2) \big ) \cr
&{}- 144 \, b (1 -  (-1)^\ta) (\ell+1 ) ( \ell + 2t +2) \, , \cr
a_{20,t\ell} ={} & {\ts {2\over 3}}(\ta-2)(\ta-1)(\ta+2)(\ta+3)(\ell+1)
(\ell+2\ta+2)\cr
\noalign{\vskip-4pt}
& \hskip 4cm {}\times (\ell+\ta-1)(\ell+\ta)(\ell+\ta+3)(\ell+\ta+4) \cr
&{}- 6 \, a \big ( (1+ (-1)^\ta) (t-2) (t+3)(\ell + t ) ( \ell+t+3)
((\ell+1)(\ell+2t+2) -4  ) \cr
\noalign{\vskip-4pt}
& \hskip 1cm {} + (1 -  (-1)^\ta) (t-1)(t+2) (\ell + t - 1) ( \ell+t+4)
((\ell+1)(\ell+2t+2) + 4 ) \big )\cr
&{}+ 48 \, b \big (2 (1+ (-1)^\ta)(\ell + 1 ) ( \ell+2t+2) \cr
\noalign{\vskip-4pt}
& \hskip 3cm {} - (-1)^\ta (t-1) (t+2)(\ell + t - 1 ) ( \ell+t+4) \big ) 
\, ,\cr
a_{11,t\ell}  = {}& 2(\ta-3)\ta(\ta+1)(\ta+4)(\ell+1)(\ell+2\ta+2) \cr
\noalign{\vskip-4pt}
& \hskip 4cm {}\times (\ell+\ta-2)(\ell+\ta+1)(\ell+\ta+2)(\ell+\ta+5) \cr
& +12\, a \big ( t(t+1) ( 11(\ell+1)(\ell+2t+2)(\ell+t+1)(\ell+t+2) 
+12(-1)^t
(t-3)(t+4) ) \cr
\noalign{\vskip-4pt}
& \hskip 1.2cm {}-  6(1-(-1)^t) 
(\ell+1)(\ell+2t+2)((\ell+t+1)(\ell+t+2)+(t-3)(t+4)
\big ) \cr
&{}+ 48 \, b \big ( 6 (1+ (-1)^\ta)(\ell + 1 ) ( \ell+ 2t + 2) \cr
\noalign{\vskip-4pt}
& \hskip 3cm {} + (-1)^\ta (t-3) (t+4)(\ell + t +1 ) ( \ell+t+2)  \big ) 
\cr
&{}+ 480\, c \big ( (1+(-1)^\ta) (\ell+1)(\ell+2\ta+2) - 2(-1)^\ta
(\ell+\ta+1)(\ell+\ta+2) \big ) \, , \cr
a_{10,t\ell}  = {}& {\ts {5\over 3}} 
(\ta-3)(\ta-1)(\ta+2)(\ta+4)(\ell+1)
(\ell+2\ta+2)\cr
\noalign{\vskip-4pt}
& \hskip 4cm {}\times (\ell+\ta-2)(\ell+\ta)(\ell+\ta+3)(\ell+\ta+5) \cr
&{} + 60\, a (\ell+1)(\ell+2t+2)(\ell+t)(\ell+t+3)\big(
2(t-1)(t+2) + 1 +  (-1)^\ta \big ) \cr
&{}+ 60 \,  \big (a(t-3)(t+4)  - 4 b - 8c \big )  (1 +  (-1)^\ta) 
(\ell+1 ) ( \ell + 2t +2) \, ,
\cr a_{00,t\ell} ={} & 
(\ta-3)(\ta-2)(\ta+3)(\ta+4)(\ell+1)(\ell+2\ta+2)\cr
\noalign{\vskip-4pt}
& \hskip 4cm {}\times (\ell+\ta-2)(\ell+\ta-1)(\ell+\ta+4)(\ell+\ta+5) \cr
& +12\, a \big ( (\ell+1)(\ell+2t+2)(\ell+t-1)(\ell+t+4)(11(t-2)(t+3) -3 +
3(-1)^t ) \cr
\noalign{\vskip-4pt}
& \hskip 1.2cm {}-  (1-(-1)^t) (\ell+1)(\ell+2t+2)(t-3)(t+4) \cr
\noalign{\vskip-4pt}
& \hskip 1.2cm {} + 6 (-1)^t(t-3)(t-2)(t+3)(t+4) \big ) \cr
&{}+ 48\, b \big ( 63 (\ell+1)(\ell+2t+2) \cr
\noalign{\vskip-4pt}
& \hskip 1.2cm {} + (-1)^t (
(\ell+t+1)(\ell+t+2)(t^2+t-9) - 9(t-1)(t+2) + 54 ) \big ) \cr
&{} + 480\, c \big ( (1+(-1)^\ta) (\ell+1)(\ell+2\ta+2) - 2(-1)^\ta
((\ell+\ta+1)(\ell+\ta+2) - 9 )  \big ) \, . & \afour
\cr}
$$
Expanding $\hf$ from \ffour\ gives
\eqnn\bfour
$$\eqalignno{
\!\!\!\!\!\! b_{3,\ell} ={}&  - {1\over 5}\, 2^{\ell-2} {(\ell+1)!^2
\over (2\ell+2)!}\,(\ell+3)\ell 
\Big ( (\ell+2)(\ell+1)\big ( {\ts {1\over 9}} (\ell+4)(\ell-1) + a \big )
- 4(a-b) \Big )  \, , \cr
\!\!\!\!\!\! b_{2,\ell} ={}& - {1\over 3} 2^{\ell-2} {(\ell+1)!^2
\over (2\ell+2)!}\, (\ell+2)(\ell+1)\Big ( (\ell+3)\ell\, \big ( 
{{\ts{1\over 3}}}(\ell+5)(\ell-2) - a \big ) - 4(3a + b +2c) \Big ) \, ,  \cr
\!\!\!\!\!\! b_{1,\ell} ={}&  - {1\over 5}\, 2^{\ell} {(\ell+1)!^2
\over (2\ell+2)!}\,\Big( \quar (\ell+4)(\ell+2)(\ell+1)(\ell-1) 
\big ( (\ell+5)(\ell-2) - 11a \big ) \cr
\noalign{\vskip-4pt}
&\hskip 3.4cm {} + 9b (\ell+2)(\ell+1) + 6 (3a+2b+5c) \Big ) \, , \cr
\!\!\!\!\!\! b_{0,\ell} ={}&  - {1\over 3}\, 2^{\ell} {(\ell+1)!^2
\over (2\ell+2)!}\,\Big( {\ts {1\over 12}}
(\ell+5)(\ell+4)(\ell+3)\ell(\ell-1)(\ell-2) 
\cr
\noalign{\vskip-4pt}
&\hskip 3.4cm {} -{\ts {5\over 4}}(\ell+3)(\ell+2)(\ell+1) \ell \, a \cr
\noalign{\vskip-2pt}
&\hskip 3.4cm {} + (\ell+2)(\ell+1)(9a+7b+2c) -18  (2a+b+c) \Big ) \, . 
& \bfour \cr}
$$
The results \afour\ and \bfour\ satisfy the six relations required by \bA\ 
when
$p=4$ and furthermore we have $a_{10,1\ell}=a_{11,0\ell}=a_{20,1\ell}=
a_{20,2\ell}=a_{21,0\ell}=a_{21,2\ell}=a_{22,0\ell}=a_{22,1\ell}=0$
in accord with \UA.

For the $p=4$ case the conformal partial wave expansion may be rewritten
in the form
\eqnn\exfour$$\eqalignno{ \!\!\!
a_{n'm'} = {}& \delta_{n'0}  \delta_{m'0} + 
\sum_{n=1}^4 C_n \, a_{n'm'}\big (\B_{nn}\big ) 
+ \sum_{n=0}^2 C_{n{+2}\, n} \, a_{n'm'}\big (\B_{n{+2} \, n}\big )  +  
C_{40} \, a_{n'm'}\big (\B_{40}\big ) \cr 
&{} + \sum_{{0\le m \le n\le 2 \atop \ta,\ell}}
\!\!\! {\hat A}_{nm,t\ell} \, a_{n'm'}\big 
(\A^{2\ta+\ell}_{nm,\ell}\big ) \cr
\noalign{\vskip -2pt}
& {}+ {\ts {1\over 140}} \!\! \sum_{\ell = 0,2,\dots}\!\!\! \Big ( 
3 A_{22,2\,\ell{+2} } \, a_{n'm'}\big (\C_{33,\ell}\big ) 
-  \half A_{11,1\,\ell{+4} } \, a_{n'm'}\big (\C_{32,\ell{+1}}\big ) \cr
\noalign{\vskip -8pt}
& \hskip 2.5cm {}+ 
{\ts {1\over 8}} A_{00,0\,\ell{+4} } \, a_{n'm'}\big (\C_{31,\ell}\big ) 
-{\ts {1\over 8}} b_{0,\ell{+5} } \, a_{n'm'}\big (\C_{30 ,\ell{+1}}\big ) 
\Big ) \, , & \exfour \cr}
$$
where we have
\eqn\Csf{
C_{42} = - {\ts {1\over 490}}\, A_{11,12} \, , \qquad
C_{40} = - {\ts {1\over 1960}}\, b_{0,3} \, ,
}
and for $t=3$ we define
\eqn\Att{\eqalign{
{\hat A}_{20,3\ell} = {}& A_{20,3\ell} + {\ts {1\over 160}} \, 
b_{0,\ell{+3}}
\, , \qquad
{\hat A}_{21,3\ell} = A_{21,3\ell} - {\ts {1\over 160}} \, 
A_{00,0\,\ell+3}
\, , \cr
{\hat A}_{22,3\ell} = {}& A_{22,3\ell} + {\ts {1\over 40}} \, 
A_{11,1\,\ell+2}\, .\cr}
}
{}From \afour\ and \bfour\ using \Aha\ we get for the $p=4$ case
\eqn\Afzfour{
{\hat A}_{00,1\ell} = 2^{\ell+1} {(\ell+2)!^2\over (2\ell+4)!} \, a 
\, ,
}
and from \Aoo
\eqn\Affour{\eqalign{
{\hat A}_{10,2\ell} = {}& 2^{\ell}\, {(\ell+3)!^2\over 3(2\ell+6)!}
\big ( (\ell+1)(\ell+6) b - 2a + 4b - c\big )\, , \cr
{\hat A}_{11,2\ell} = {}& 2^{\ell} {(\ell+3)!^2\over 3(2\ell+6)!}
\big ( (\ell+3)(\ell+4) b + c \big ) \, .
\cr}
}
For the remaining twist 4 contribution we have
\eqn\Afop{
{A}_{00,2\ell} = 2^{\ell} {(\ell+3)!^2\over 3(2\ell+6)!}
\big ( (\ell+1)(\ell+6) b + c \big ) \, .
}
Furthermore \Att\ gives for twist 6
\eqnn\Ahft
$$\eqalignno{
{\hat A}_{20,3\ell} = {}&
2^{\ell-2}\, {(\ell+4)!^2\over 15(2\ell+8)!}
\big ( (\ell+3)(\ell+4)(\ell+5)(\ell+6) a \cr
\noalign{\vskip -6pt}
& \hskip 3.3cm {} - 2(\ell+4)(\ell+5)(3a+3b+c) +6(4a+2b+3c) \big )\, , \cr
{\hat A}_{21,3\ell} = {}&
2^{\ell-3}\, {(\ell+4)!^2\over 5(2\ell+8)!}
\big ( (\ell+2)(\ell+4)(\ell+5)(\ell+7) a 
- 4(\ell+4)(\ell+5)b - 12c \big )\, , \cr
{\hat A}_{22,3\ell} = {}&
2^{\ell-3}\, {(\ell+4)!^2\over 15(2\ell+8)!}(\ell+4)(\ell+5)
\big ( (\ell+3)(\ell+6) a + 4 c \big )\, , & \Ahft \cr}
$$
and we also have
\eqn\Ahff
{\eqalign{
{A}_{10,3\ell} = {}&
2^{\ell-3}\, {(\ell+4)!^2\over 3(2\ell+8)!} \,
(\ell+1)(\ell+3)(\ell+6)(\ell+8) a  \, , \cr
{A}_{11,3\ell} = {}&
2^{\ell-2}\, {(\ell+4)!^2\over 15(2\ell+8)!}(\ell+4)(\ell+5)
\big ( 3(\ell+1)(\ell+8) a + 2c \big )\, , \cr
{A}_{00,3\ell} = {}&
2^{\ell-3}\, {(\ell+4)!^2\over 5(2\ell+8)!}
\big ((\ell+1)(\ell+2)(\ell+7)(\ell+8) a \cr
\noalign{\vskip -6pt}
& \hskip 3.3cm {} + 4 (\ell+1)(\ell+8) (b + {\ts {1\over 3}} c) 
+ 4c \big )\, . \cr}
}
Using \Ck\ and \Cn\ gives $C=1$ again as required and for the coefficients 
for the contributions of $\half$-BPS operators
\eqn\Cfour{
C_1 = {\ts{1\over 3}}a \, , \quad
C_2 = {\ts{1\over 60}}(2b+c) \, , \quad C_3 = {\ts{1\over 350}} ( a + c) 
\, ,\quad 
C_4 = {\ts{1\over 2^2.3^2.5.7^2}} (2+ 2a + b) \, .
}
{}From \Cs, \Cst\ and \Csf\ we obtain
\eqn\Cres{\eqalign{
C_{20}= {}& {\ts {1\over 54}} (4b-2a-c) \, , \hskip 2.3cm
C_{31}= {\ts {9\over 2000}} (6a-4b-c) \, , \cr
C_{42}= {}& {\ts {1\over 3.5.7^3}} (24-15a-2b-4c) \, , \qquad
C_{40}= {\ts {1\over 3.5^2.7^3}} (84-153a+61b+11c) \, . \cr}
}

To summarise some of the above results we first note from \Ctwo, \Cthree\ 
and \Cfour\ that in each case $C_1 = {\ts{1\over 3}}a$. The corresponding
short multiplet $\B_{11}$ is special in that it contains the energy 
momentum tensor as well as the $SU(4)$ current. The contributions of these 
operators in the operator product expansion are constrained by Ward 
identities \DO. 
For the energy momentum tensor the coefficient of its contribution in the 
operator product expansion in four dimensions to the four point function,
by applying  \apart\ and \CT\ in this case,
is $16 p^2/9C_T$, where $C_T/S_4{\!}^2$ is the coefficient of the 
energy momentum tensor two point function. In the expansion of
$a_{00}(\B_{11})$ the contribution corresponding the energy momentum
tensor is ${2\over 15} u G^{(2)}_4 (u,v)$ so that this requires
$C_1 = 40p^2/3C_T$, where with our normalisations $C_T=40(N^2-1)$
(where $40 = 6 \times {4\over 3} + 4 \times 4 + 16$ reflecting the
decomposition into the contributions of the scalars, fermions and vector
in the elementary $\N=4$ multiplet).
Assuming that for arbitrary $p$ only $S_{p(1,0)}$ in $\G_0$ gives a 
non zero $C_1$, as we have found for $p=2,3,4$ and as may be expected since
this is the only contribution involving two particle reducible graphs, then 
its coefficient must in general be
\eqn\agen{
a= {p^2 \over N^2-1} \, .
}
This result was found exactly \hpert\ for $p=4$ even allowing for an admixture 
of double trace operators in the $\half$-BPS operators whose correlation
function is being considered. In addition we note that the expansion coefficients 
for long multiplets with twist $< 2p$ are suppressed in the large $N$ limit
so that there are only $1/N^2$ contributions. The results for 
${\hat A}_{00,1\ell}$ given by  \Afztwo, \Afzthree\ and \Afzfour\ are 
identical in form. This also applies to 
${\hat A}_{10,2\ell}, \, {\hat A}_{11,2\ell}$,
as shown by  \Atoo\ and  \Affour, and $A_{00,2\ell}$,
as given by \Atop\ and \Afop, so long as we use the large $N$ relations
$b=2a$ from \athreeo\ for $p=3$ and $b=a, \, c=2a$ from \afouro\ for 
$p=4$.
This result, which plays a crucial role in our subsequent discussions 
using perturbation theory and also results from large $N$ obtained via AdS/CFT,
is dependent on there being only contributions for semi-short multiplets $\C_{nm}$ 
in the superconformal partial wave expansions, as in  \extwo, \exthree\ or 
\exfour, for $n=p-1$, with others decoupled for large $N$. In 
the large $N$ limit we may also note that the coefficients for $\half$-BPS
multiplets $\B_{nn}$ are suppressed except when $n=p$ and the $\quar$-BPS
multiplets $\B_{n{+2}\, n}$, from \Cqq\ and  \Cres, are absent for 
$n=0, \dots p-3$.

\newsec{Large $N$, Strong Coupling Results}

The dynamical contributions to the $\ha$-BPS four-point functions is 
contained solely in the function $\H$ which is constrained by crossing symmetry
\eqn\cross{
\H(u,v;\sigma,\tau) = {1\over v^2}\, \H(u/v,1/v;\tau,\sigma) =
\Big ( {u\over v} \Big )^p \tau^{p-2}\, \H (v,u;\sigma/\tau,1/\tau) \, .
}
The results obtained via the  AdS/CFT correspondence, for the large $N$ limit, are 
expressible in terms of  conformal four-point integrals which may be reduced to the 
two variable functions $\oD_{\De_1\De_2\De_3\De_4}(u,v)$ for arbitrary $\De_i$. 
The properties of these functions have been explored in detail \refs{\Freed,\DO} 
and they satisfy various symmetry and other relations, some of which are listed in
appendix C.

For $p=2$ we have 
\eqn\Htwo{
\H(u,v;\sigma,\tau) = - {4\over N^2} \, u^2 \oD_{2422}(u,v) \, ,
}
and for $p=3$ we have \ADHS
\eqn\Hthree{
\H(u,v;\sigma,\tau) = - {9\over N^2} \, u^3 \big ( ( 1+\sigma+\tau) 
\oD_{3533}
+ \oD_{3522} + \sigma \, \oD_{2523} + \tau \, \oD_{2532} \big ) \, ,
}
while for $p=4$ we have \Degen
\eqnn\Hfour
$$\eqalignno{
\H(u,v;\sigma,\tau) = - {4\over N^2} \, u^4 & \Big ( (1+\sigma^2 + \tau^2
+ 4 \sigma + 4 \tau + 4 \, \sigma\tau ) \oD_{4644} \cr 
&{}+ 2 ( \oD_{4633} + \oD_{4622} ) + 2\sigma^2 ( \oD_{3634} + \oD_{2624} )
+2 \tau^2 ( \oD_{3643} + \oD_{2642} ) \cr
&{} - 4\, \sigma( \oD_{4624} -2 \oD_{3623} ) - 4 \, 
\tau( \oD_{4642} -2 \oD_{3632} ) \cr
&{} - 4\, \sigma\tau( \oD_{2644} -2 \oD_{2633} ) \Big ) \, . & \Hfour \cr}
$$
In each case the results have been rewritten to ensure that the $\oD$ functions
are multiplied by the maximum
overall power of $u$. The crossing relations \cross\ also follow straightforwardly
using symmetry relations for $\oD$ functions listed in appendix C.

Defining
\eqn\defs{
s = \half ( \De_1 + \De_2 - \De_3 - \De_4 ) \, ,
}
then for $s=0,1,2,\dots$ $\oD_{\De_1 \De_2 \De_3 \De_4}(u,v)$ can be written in 
the form $\ln u f(u,v) + g(u,v) $ where $f(u,v),g(u,v)$ have power 
series expansions in powers of $u$ and $1-v$, although for $g$ it is
necessary to allow negative powers $u^{-s+m}, \, m=0,1,2\dots$. The case
for $s=-1,-2,\dots$ may also be accommodated by virtue of
\eqn\relDD{
\oD_{\De_1 \De_2 \De_3 \De_4}(u,v) = u^{-s} 
\oD_{\De_4 \De_3 \De_2 \De_1}(u,v) \, .
}
The $\ln u$ terms of
course lead to contributions to anomalous dimensions of order $1/N^2$. From
\Htwo, \Hthree\ and \Hfour\ the operators which gain anomalous dimensions
must have a twist of at least $2p$.

However the potentially singular contributions present in $g(u,v)$
involving negative powers of $u$ play a significant role. For
$s$ a positive integer the $\oD$ function can be decomposed in the form
\eqn\deDD{
\oD_{\De_1 \De_2 \De_3 \De_4}(u,v) = \oD_{\De_1 \De_2 \De_3 \De_4}(u,v)_{\rm reg.}
+ \oD_{\De_1 \De_2 \De_3 \De_4}(u,v)_{\rm sing.} \, ,
}
where the first regular part has and expansion involving $u^m$ and $u^m\ln u$ for
$m=0,1,\dots$ and
\eqn\Dsing{\eqalign{
\oD_{\De_1 \De_2 \De_3 \De_4}(u,v)_{\rm sing.} = 
{} & u^{-s} \, 
{\Gamma(\De_1-s)\Gamma(\De_2-s)\Gamma(\De_3)\Gamma(\De_4)\over 
\Gamma(\De_3 +\De_4) } \cr
& \times
\sum_{m=0}^{s-1} (-1)^m (s-m-1)! \, 
{(\De_1-s)_m (\De_2-s)_m (\De_3)_m (\De_4)_m\over m! \, 
(\De_3+\De_4)_{2m} } \cr
\noalign{\vskip -6pt}
&\qquad \ {} \times u^m F(\De_2-s+m,\De_3+m;\De_3+\De_4+2m;1-v) \, .\cr}
}
In \Htwo\ the $\oD$ function has $s=1$ while in \Hthree\ we have  $s=1,2$
and in \Hfour\ $s=0,1,2,3$. In general with the $SU(4)$ decomposition
given by \exH\ we may write
\eqn\AB{
A_{nm}(u,v) = B_{nm}(u,v)+ {\rm O}(u^p\ln u, u^p) \, ,
}
where $B_{nm}$ is calculated by using \Dsing. Since there are no $\ln u$
terms in $B_{mn}(u,v)$ there are no anomalous dimensions obtained in the conformal 
wave expansion which involve operators with twist $2 \le \De-\ell < 2p$.
It is the purpose here to show that these contributions cancel exactly the
corresponding long supermultiplet contributions obtained by the free-field
calculations for large $N$ in section 3. 
This indicates that the corresponding long supermultiplets 
$\A^{\Delta}_{nm,\ell},\,\,0<\Delta-\ell<2p$,
decouple from the spectrum in the large $N$ limit. As mentioned earlier, this
depends on requiring that the contributions from semi-short multiplets 
$\C_{nm,\ell}$ in the partial wave expansion 
are solely those for $n=p-1$, with others disappearing for large $N$.

For $p=2$ it is easy to see that
\eqn\peqtwoterms{
B_{00}(u,v)=-{4\over 3 N^2}\,u\,F(3,2;4;1-v)\,.
}
For $p=3$ from \Hthree\ we get, with the aid of various $\oD$ identities,
\eqn\Athree{\eqalign{
A_{11}(u,v) = {}& - {3\over 2N^2}u^3 \Big ( 2 \oD_{3533}(u,v) - \oD_{1533}(u,v)
+ {1\over uv^2} \Big ) \, , \cr
A_{10}(u,v) = {}& - {3\over 2N^2}u^3 \Big ( \oD_{2523}(u,v) - \oD_{2532}(u,v)
\Big ) \, , \cr
A_{00}(u,v) = {}& - {3\over 2N^2} u^3 \Big ( 8 \oD_{3533}(u,v) + 6 \oD_{3522}(u,v)
- \oD_{1533}(u,v) + {1\over uv^2} \Big ) \, , \cr}
}
which leads to
\eqn\peqthreeterms{\eqalign{
B_{11}(u,v)& =-{3\over 5 N^2}\,u^2\,\left(F(4,3;6;1-v)
+ {5 \over 2\,v^2}\right)\,,\cr
B_{10}(u,v)& =
 {3\over 5 N^2}\,u^2\,(1-v)\,F(5,3;6;1-v)\,,\cr
B_{00}(u,v)& = -{3\over 5 N^2}\,u\,\left(5 F(3,2;4;1-v)+ u\,
 F(4,3;6;1-v)+{5 \, u\over 2\, v^2}\right)\,,}
}
using \Dsing\ and for $B_{10}$ a result from appendix C. It is also useful to note
that $v^{-2} = F(4,2;4;1-v)$. 
In a similar fashion for $p=4$ we may obtain
\eqnn\Afour
$$\eqalignno{
A_{22} = {}& - {2 \over 5N^2}u^4 \Big ( \oD_{4644} +\oD_{2633} - \oD_{2644} -{\ts{1\over 3}}
\big ( \oD_{1634} + \oD_{1643} \big ) + {2\over 3} \, {1\over uv^3}(1+v) \Big ) \, ,
\cr A_{20} = {}& - {4 \over 5N^2}u^4 \Big (  - \oD_{2633}  -{\ts{1\over 3}}
\big ( \oD_{1634} + \oD_{1643} \big ) + {2\over 3} \, {1\over uv^3}(1+v) \Big ) \, , \cr
A_{11} = {}& - {4 \over 5N^2}u^4 \Big ( 8 \oD_{4644} + {\ts {20\over 3}} \oD_{4633}  - 
{\ts{1\over 3}} \oD_{2633} - {\ts{14\over 3}} \oD_{2644}  -
\big ( \oD_{1634} + \oD_{1643} \big ) \cr
\noalign{\vskip - 4pt}
&\qquad\qquad\quad {} + 2 \, {1\over uv^3}(1+v) +{10\over 3}\, {1\over u^2v^2} 
\Big ) \, , \cr
A_{00} = {}& - {4 \over N^2}u^4 \Big ({\ts{5\over 2}} \oD_{4644} + {\ts {10\over 3}} \oD_{4633}  
+ 2 \oD_{4622} +  {\ts{1\over 30}} \oD_{2633} - {\ts{5\over 6}} \oD_{2644}  - {\ts{1\over 10}}
\big ( \oD_{1634} + \oD_{1643} \big ) \cr
\noalign{\vskip - 4pt}
&\qquad\qquad\quad {} + {1\over 5} \, {1\over uv^3}(1+v) +{2\over 3}\, {1\over u^2v^2}
\Big ) \, , & \Afour \cr}
$$
and also
\eqn\Afoura{\eqalign{
A_{21} = {}& - {2 \over 5N^2}u^4 \Big ( \oD_{3634} - \oD_{3643} 
- \oD_{1634} + \oD_{1643}  -2 \, {1\over uv^3}(1-v) \Big ) \, , \cr
A_{10} = {}& - {2 \over 3N^2}u^4 \Big (  4 \big (\oD_{3623}  - \oD_{3632} \big ) +5
\big ( \oD_{3634} - \oD_{3643} \big ) \cr
\noalign{\vskip - 4pt}
&\qquad\qquad\qquad {} - \oD_{1634} + \oD_{1643}  -
2 \, {1\over uv^3}(1-v) \Big ) \, . \cr}
}
In \Afour\ $\oD_{1634} + \oD_{1643}$ may be further simplified using a result in 
appendix C. Using \Dsing\ and an extension from appendix C this leads to
\eqn\peqfourterms{\eqalign{
B_{22}(u,v)& = - {4\over 5 N^2}\,u^3\,\left({\ts{2\over 5}}F(5,3;6;1-v)+
{\ts{6 \over 35}}
F(5,4;8;1-v)+{\ts{1\over 3}}(1+v){1\over v^3}\right)\,,\cr
B_{21}(u,v)& = {4\over 5 N^2}\,u^3\,(1-v)\left({\ts{1\over
7}}F(6,4;8;1-v)+{1\over v^3}\right)\,,\cr
B_{20}(u,v)& = -{8\over 5 N^2}\,u^3\,\left(-{\ts{2\over 5}}F(5,3;6;1-v)
+{\ts{1\over 3}}(1+v){1\over v^3}\right)\,,\cr
B_{11}(u,v)& = -{8\over 5 N^2}\,u^2\,\Big({\ts{2\over 3}}F(4,3;6;1-v)
+{\ts{5\over 3}}\, {1\over v^2}-{\ts{2\over 15}}\,u\,F(5,3;6;1-v)\cr
&\qquad \qquad\quad \qquad  +{\ts{8\over 35}}\,u\,
F(5,4;8;1-v)+(1+v){u\over v^3}\Big)\,,\cr
B_{10}(u,v)& ={4\over 5 N^2}\,u^2\,(1-v)\left({\ts{4\over 3}}F(5,3;6;1-v)+
{\ts {5\over 21}}\,u\,F(6,4;8;1-v)+{\ts{5\over 3}}\, {u\over v^3}
\right)\,,\cr
B_{00}(u,v)& = -{8\over 5 N^2}\,u\,\Big({\ts{10\over 3}}F(3,2;4;1-v)+
{\ts{2\over 3}}\,u\,F(4,3;6;1-v)+{\ts{5\over 3}}{u\over v^2}\cr
&\qquad\qquad \quad\quad  +{\ts{1\over 15}}
\,u^2\,F(5,3;6;1-v)+{\ts{1\over 7}}\,u^2\,
F(5,4;8;1-v)+\half (1+v){u^2\over v^3}\Big)\,, \cr }
}
where also $v^{-3} = F(6,3;6;1-v)$.

The conformal partial wave expansion
\eqn\confpwe{
B_{nm}(u,v)=\sum_{t,\ell}B_{nm,t\ell}\,u^t\,G^{(\ell)}_{\ell+2t+4}(u,v)\,,
}
would involve contributions for operators which have no anomalous dimensions.
To analyse these we need to compute $B_{nm,t\ell},\,1\leq t\leq p-1$ for 
$p=2,3,4$. Since
\eqn\lowt{
G^{(\ell)}_{\ell+2t}(0,v) = g_{t,\ell}(1-v) \, ,
}
where $g_{t,\ell}$ is defined in \gell,
then for the lowest twist contributions for each $B_{nm}$ it is sufficient 
to use, since starting from $\oD_{n_1n_2nn}$
only hypergeometric functions of this form  appear in \peqtwoterms,
\peqthreeterms\ and \peqfourterms,
\eqn\exF{
F(a,b;2b;x) = \sum_{\ell=0,2,\dots} r_{ab,\ell} \, g_{a,\ell}(x) \, , 
}
where, as demonstrated in appendix D,
\eqn\cellab{
r_{ab,\ell}={1\over 2^\ell\,({1\over 2}\ell)!}\,{(a)_\ell 
(a-b)_{{1\over 2}\ell}\over(a+{1\over 2}\ell-{1\over 2})_{{1\over 2}\ell}
(b+{1\over 2})_{{1\over 2}\ell}}\,.
}
More generally for application to the required expansion of $B_{nm}$ this is 
extended in appendix D to an expansion in conformal partial waves with 
increasing twist
\eqn\expo{
u^{a-2}F(a,b;2b;1-v)=\sum_{{\ell=0,2,\dots\atop j=0,1,\dots}}
c_{ab;j,\ell}\,u^{a-2+j} G^{(\ell)}_{\ell+2a+2j}(u,v)\,,
} 
where the first few cases of $c_{ab;j,\ell}$ are
\eqn\solto{\eqalign{
& c_{ab;0,\ell} =r_{ab,\ell}\,,\cr
& c_{ab;1,\ell}= {a b (a-2b)\over 2(4b^2-1)} r_{a+1\,b+1,\ell}
-{\ts{1\over 4}}r_{a-1\, b-1,\ell+2}\,,\cr
& c_{ab;2,\ell} = a(a+1)(b+1){(a-2b-1)(a-2b)\over 16 (2b+1)^2(2b+3)}
\,r_{a+2\,b+2,\ell}-a\,{a(a-3b-1)+b\over 16(2a-1)(2b+1)}\,r_{ab,\ell+2}\,.
}}

For $p=2$ we may then find for the partial wave expansion coefficients
\eqn\petwo{
B_{00,1\ell}= - {4\over 3N^2}\, c_{3\,2;0,\ell}  
= -{1\over N^2}2^{\ell+3}{(\ell+2)!^2\over (2\ell+4)!}\,,
}
which clearly cancels $\hat{A}_{00,1\ell}$ in \Afztwo\ for $a$ as in
\atwo.
For $p=3$ we find that, using for $B_{10}$ a recurrence relation for
$(1-v)G^{(\ell)}_\Delta (u,v)$,
\eqnn\findpthree
$$\eqalignno{
B_{11,2\ell}& =  -{3\over 5N^2} \big ( c_{4\,2;0,\ell} + {\ts {5\over 2}}
c_{4\,2;0,\ell} \big )  = -{3\over N^2}\, 2^{\ell}
{(\ell+3)!\over (2\ell+6)!} (\ell^2+7\ell+14)\,,\cr
B_{10,2\ell}& =  -{6\over 5N^2} \, c_{5\,3;0,\ell-1} 
=-{3\over N^2}\, 2^{\ell}{(\ell+3)!^2\over (2\ell+6)!}(\ell+1)(\ell+6)\,,\cr
B_{00,2\ell}& = - {3\over N^2}\big ( c_{3\,2;1,\ell} + {\ts {1\over 5}}
c_{4\,3;0,\ell} + \half c_{4\,2;0,\ell} \big ) 
=-{3\over N^2}\, 2^{\ell}{(\ell+3)!^2\over (2\ell+6)!}(\ell^2+7\ell+8)\,,\cr
B_{00,1\ell}& =  - {3\over N^2}\, c_{3\,2;0,\ell} 
= -{9\over N^2}\, 2^{\ell+1}{(\ell+2)!^2\over (2\ell+4)!}\,, & \findpthree \cr}
$$
which cancel the corresponding free-field cases in \Afzthree, \Aoo,
\Atoo\ and \Atop\ with $a,b$ as in \athreeo.
For $p=4$ in a similar fashion we find that
\eqn\findpfourthree{\eqalign{
B_{22,3\ell}& =-{1\over N^2}\, 2^{\ell+1}{(\ell+4)!^2\over 
15(2\ell+8)!}(\ell+4)(\ell+5)(\ell^2+9\ell+26)\,,\cr
B_{21,3\ell}& =-{1\over N^2}\, 2^{\ell+1}{(\ell+4)!^2\over 
5(2\ell+8)!}(\ell+1)(\ell+8)(\ell^2+9\ell+22)\,,\cr
B_{20,3\ell}& =-{1\over N^2}\, 2^{\ell+2}{(\ell+4)!^2\over 
15(2\ell+8)!}(\ell+1)(\ell+2)(\ell+7)(\ell+8)\,,\cr
B_{11,3\ell}& =-{1\over N^2}\, 2^{\ell+2}{(\ell+4)!^2\over
15(2\ell+8)!}(\ell+4)(\ell+5)(3\ell^2+27\ell+28)\,,\cr
B_{10,3\ell}& =-{1\over N^2}\, 2^{\ell+1} {(\ell+4)!^2\over 3(2\ell+8)!}
(\ell+1)(\ell+3)(\ell+6)(\ell+8)\,,\cr
B_{00,3\ell}& =-{1\over N^2}\, 2^{\ell+1}{(\ell+4)!^2\over
15(2\ell+8)!}(3\ell^2+27\ell+26)\,,\cr
B_{11,2\ell}& =-{1\over N^2}\, 2^{\ell+4}{(\ell+3)!^2\over 3
(2\ell+6)!} (\ell^2+7\ell+14)\,,\cr
B_{10,2\ell}& =-{1\over N^2}\, 2^{\ell+4}{(\ell+3)!^2\over 3
(2\ell+6)!}(\ell+1)(\ell+6)\,,\cr
B_{00,2\ell}& =-{1\over N^2}\, 2^{\ell+4}{(\ell+3)!^2\over 3
(2\ell+6)!}(\ell^2+7\ell+8)\,,\cr
B_{00,1\ell}& = -{1\over N^2}\, 2^{\ell+5}{(\ell+2)!^2\over
(2\ell+4)!}\,,}
}
which cancel exactly the corresponding free-field cases in \Afzfour,
\Affour, \Afop, \Ahft\ and \Ahff\ for $a,b,c$ as in \afouro.

These cancellations are very non trivial and provide a strong
consistency check on the results \Htwo, \Hthree\ and \Hfour. The
coefficients of $\ln u$ in the large $N$ results for $A_{nm}(u,v)$
may also be expanded to give results for large $N$ anomalous
dimensions, for $p=2$ and $p=3$ these were quoted in \SCFT\ and
\ADHS\ respectively, for $p=4$ see \Nirschl.

\newsec{Perturbation Theory Results}

In order to express the results from perturbation theory it is convenient
to write solutions of the crossing symmetry relations \cross\ in 
terms of invariant functions of $u,v$.
For $p=2$ 
\eqn\ptwo{
\H(u,v;\sigma,\tau) = {u\over v} \, \F(u,v) \, , 
}
while for $p=3$
\eqn\pthree{
\H(u,v;\sigma,\tau) = {u\over v} \, \F(u,v) + {u^2 \over v^2} \Big (
\sigma \, \F(1/v,u/v) + \tau \, \F(v,u) \Big ) \, ,
}
and for $p=4$ there are two functions $\F,\tF$.
\eqn\pfour{\eqalign{
\H(u,v;\sigma,\tau) = {}& {u\over v} \, \F(u,v) + {u^2 \over v} \Big (
\sigma \, \tF(v,u) + \tau \, {1\over v^2}\tF(1/v,u/v) \Big )  \cr
&{} + {u^3 \over v^2} \Big ( \sigma^2  \, \F(1/v,u/v) +
\tau^2 \, {1\over v}\F(v,u) + \sigma \tau \, \tF(u,v) \Big ) \, . \cr}
}
In each case we must have to satisfy \cross
\eqn\crossF{
\F(u,v)= {1\over  v} \F(u/v,1/v) \, , \qquad 
\tF(u,v)= {1\over v}\tF(u/v,1/v) \, ,
}
while for $p=2$ we have in addition
\eqn\crt{
\F(u,v)=  \F(v,u) \, .
}

To first order in perturbation theory there is a simple general formula, for 
$\lambda = g_{YM}^2N/4\pi^2$,
\eqn\onel{
\F_1(u,v) = \tF_1(u,v) = - {p^2\lambda \over 2 N^2} \, \Phi^{(1)}(u,v) \, .
}
If we define
\eqn\IJp{\eqalign{
\I(u,v) = {}& \quar (1+v) \Phi^{(1)}(u,v)^2
+ \Phi^{(2)}(u,v) + {1\over v} \Phi^{(2)}(u/v,1/v)  \, , \cr
\J(u,v) = {}& \quar u \, \Phi^{(1)}(u,v)^2 + 
{1\over u} \Phi^{(2)}(1/u,v/u) \, , \cr}
}
then the results obtained in \pert\ and, for $p=3,4$, in \hpert\ to 
${\rm O}(\lambda^2)$ are
\eqn\FIJ{
\F_2(u,v) = {p^2\lambda^2 \over 4 N^2}\, \I (u,v) \, , \ p=3,4 \, , \quad
\tF_2(u,v) = {p^2\lambda^2 \over 4 N^2}\,  \J (u,v) \, , \ p= 4 \, ,
}
and for $p=2$
\eqn\Ftwo{
\F_2(u,v) = {\lambda^2 \over N^2} \big ( \I(u,v) + \J(u,v) \big ) \, ,
}
which is necessary to satisfy \crt.

In \onel\ and \IJp\ $\Phi^{(L)}$ are conformal loop integrals,
$\Phi^{(1)}(u,v) = \oD_{1111}(u,v)$. If
\eqn\defxx{
u = {x' \bx' \over (1-x')(1-\bx')} = x \bx \, , \quad 
v = {1 \over (1-x')(1-\bx')} = (1-x)(1-\bx) \, ,
}
with $x'$ related to $x$ as in \xx, then defining
\eqn\PPh{
{\hat \Phi}^{(L)}(x',\bx') = {\hat \Phi}^{(L)}(\bx',x') =\Phi^{(L)}(u,v) \, ,
}
allows an expression in terms of single variable polylogarithms \UD.
The form given by Isaev \Isa\ is particularly simple and for $L=1,2$
\eqn\polyL{\eqalign{
v{\hat \Phi}^{(1)}(x',\bx') = {}& - \ln x'\bx' \, \phi_1(x',\bx') +
2 \phi_2(x',\bx')\, ,
\cr 
v{\hat \Phi}^{(2)}(x',\bx') = {}& \half\ln^2\! x'\bx' \, \phi_2(x',\bx')  
-3 \ln x'\bx' \, \phi_3(x',\bx') + 6\, \phi_4(x',\bx') \, , \cr}
}
where $\phi_n$ is defined by
\eqn\phin{
\phi_n(x,\bx) = {\Li_n(x) - \Li_n(\bx) \over x - \bx } \, , \qquad
\Li_n(x) = \sum_{r=1}^\infty {x^r\over r^n} \, , \ \ {\Li_1}(x) = - \ln (1-x) \, . 
}
The $\Phi^{(L)}$ satisfy
\eqn\relPh{
\Phi^{(L)}(u,v) = \Phi^{(L)}(v,u) \ \ \Longleftrightarrow \  \
{\hat \Phi}^{(L)}(x',\bx') = {\hat \Phi}^{(L)}(1/x',1/\bx') \, ,
}
where the latter result follows from standard polylogarithm identities
for $x',\bx'<0$. In addition we have $\Phi^{(1)}(u,v) = \Phi^{(1)}(u/v,1/v)/v
= - \ln x\bx \, \phi_1(x,\bx) + 2 \phi_2(x,\bx)$.

For the perturbative analysis of the operator product expansion we consider
initially supermultiplets whose superconformal primaries are singlets
with twist 2 in free theory. In this case it is sufficient to consider
only the leading terms as $u\to 0$. For $p=2,3,4$ from
from \ptwo, \pthree, \pfour
\eqn\sing{
A_{00}(u,v)_{\rm pert.} = {u\over v} \, \F(u,v) + {\rm O}(u^2)
= A_{00,1}(u,v)  + {\rm O}(u^2) \, ,
}
and since, as shown in appendix C there are no terms in $\Phi^{(2)}(1/u,v/u)/u$
proportional to $\ln u$ for small $u$ so that $\J$ in \Ftwo\ may be neglected,
we may take
\eqn\Aone{
A_{00,1}(u,v) ={p^2 \over N^2 } \, {u\over v} \big ( - \half \lambda
\Phi^{(1)}(u,v) + \quar \lambda^2 \, \I(u,v) + {\rm O}(\lambda^3) \big )\, .
}
Using \defxx\ this may be further simplified giving
\eqn\Aoos{
A_{00,1}(u,v) 
\sim {} {p^2 \over N^2} \, u \sum_{r=1,2,\dots } \!\! \lambda^r \sum_{s=0}^r  \, 
\ln^s \! x\bx  \,  f_{rs} ( x ) \quad \hbox{as} \quad \bx \to 0 \, , 
}
where, with the above results \polyL, we obtain
\eqn\frs{\eqalign{
f_{11}(x) = {}& - {1\over 2} \, {1\over x(1-x)}\, \ln(1-x) \, ,\qquad
f_{10}(x) = - {1\over x(1-x)}\, \Li_2(x) \, , \cr
f_{22}(x) = {}& {1\over 8}\, {1\over x(1-x)} \bigg ( {1\over x}
\ln^2(1-x) + 2 \Li_2 (x) \bigg ) \, , \cr
f_{21}(x) = {}& {1\over 8}\, {1\over x(1-x)} \bigg ( - \ln^3(1-x) +
{4\over x} (1-x) \ln(1-x) \, \Li_2 (x) - 6 \big ( \Li_3(x) - \Li_3(x') \big ) 
\bigg ) \, . \cr}
}
Assuming only one operator for each $\ell$ is present with zeroth order
twist 2 we must have for $\bx\to 0$ from \exH\ and \lowt
\eqn\twist{
{\hat A}_{00}(u,v) + A_{00}(u,v)_{\rm pert.} \sim {p^2\over N^2} \, u \!
\sum_{\ell = 0,2,\dots} a_\ell \, (x\bx)^{{1\over 2}\eta_\ell}
g_{3+{1\over 2}\eta_\ell,\ell} (x) \, ,
}
where $\eta_\ell$ is the anomalous dimension for each $\ell$ and
${\hat A}_{00}(u,v)$ is obtained from free field theory with all
contributions of protected short and semi-short multiplets subtracted.
Writing
\eqn\expeta{
\eta_\ell = \lambda \, \eta_{\ell,1} + \lambda^2 \eta_{\ell,2} + \dots \, , 
\qquad a_\ell = a_{\ell,0}\big (1 + \lambda\,b_{\ell,1} + \dots \big )\, ,
}
then using \Afztwo, \Afzthree\ and \Afzfour\ for $p=2,3$ and $4$ and the
appropriate large $N$ value of $a$ we have
\eqn\azero{
a_{\ell,0} = 2^{\ell} {(\ell+1)!(\ell+2)!\over (2\ell+3)!}  \, .
}

The determination of anomalous dimensions simplifies to matching single variable 
expansions in \Aoos\ and \twist.
Using results from appendix E we first have
\eqn\exone{
f_{11}(x) = \sum_{\ell=0,2,\dots} a_{\ell,0} \, h(\ell+2) \, g_{3,\ell} (x) \, ,
}
where
\eqn\harm{
h(n)= \sum_{r=1}^n {1\over r} \, .
}
Hence we easily find that
\eqn\resone{
\eta_{\ell,1} = 2 h(\ell+2) \, ,
}
in accordance with earlier results \SCFT. Furthermore we have
\eqn\extwo{
f_{22}(x) =\half  \sum_{\ell=0,2,\dots} a_{\ell,0} \, h(\ell+2)^2 \,
g_{3,\ell} (x) \, ,
}
which demonstrates that only a single operator for each $\ell$ with twist two
at $\lambda=0$ is present. We may also write
\eqn\exonea{
f_{10}(x) = \sum_{\ell=0,2,\dots} a_{\ell,0} \big (  b_{\ell,1} \, g_{3,\ell} (x) 
+ h(\ell+2) \, g'{\!}_{3,\ell} (x) \big )  \, , \quad
g'{\!}_{t,\ell}(x) = {\pr\over \pr t} g_{t,\ell}(x) \, .
}
As shown in appendix E this gives
\eqn\aone{
b_{\ell,1} = 2h(\ell+2)^2 - 2 h(\ell+2)h(2\ell+4)
- \sum_{r=1}^{\ell+2} {1\over r^2} \, .
}
Similarly we have
\eqn\etatwo{
f_{21}(x) = \sum_{\ell=0,2,\dots} a_{\ell,0} \Big (  \big ( \half \eta_{\ell,2} 
+ b_{\ell,1}h(\ell+2)  \big ) g_{3,\ell} (x)
+ h(\ell+2)^2  g'{\!}_{3,\ell} (x) \Big )  \, , 
}
which gives according to the results of appendix E
\eqn\etares{\eqalign{
\eta_{\ell,2} = {}& -2 \sum_{r=1}^{\ell+2} {1\over r} \sum_{s=1}^r 
{(-1)^s \over s^2} - 2 h(\ell+2)
\sum_{r=1}^{\ell+2} {1\over r^2}  - 
\sum_{r=1}^{\ell+2} {1\over r^3} \big ( 1 - (-1)^r \big )  \cr
= {}& 2 \sum_{r=1}^{\ell+2} {(-1)^r\over r^2} h(r)  - 2 h(\ell+2)
\sum_{r=1}^{\ell+2} {1\over r^2}\big ( 1 + (-1)^r \big )   -
\sum_{r=1}^{\ell+2} {1\over r^3} \big ( 1 + (-1)^r \big ) \, ,\cr}
}
where we give two equivalent expressions.
We may note that $b_{0,1} = -3, \, b_{2,1} = -{1025\over 252}$ and 
$\eta_{0,2}= -3, \,\eta_{2,2} = - {925\over 216}$ which coincide with the 
revised results of Arutyunov {\it et al} \Except. For $\ell=0$ the operator 
is the Konishi scalar whose second order anomalous dimension was found in \Kon. 
The results for general $\ell$ are the same as  those obtained by very different
perturbative calculations \refs{\lipatov,\belitsky}.  The corrections to the 
coupling $b_{\ell,1}$ are universal in that they are independent of the specific 
BPS operator, or value of $p$, in accord with results in \OST.

At higher twist there are several operators for each $\ell$ which leads to
mixing effects \refs{\mixing,\Biamix}. In general we write the scale dimension 
of the superconformal primary operator belonging to the representation with
Dynkin labels $[n-m,2m,n-m]$ in a long multiplet in the form
\eqn\prim{
\Delta^I_{nm,t\ell} = 2t + \ell + \eta^I_{nm,t\ell} \, , \qquad 
t=n+1,n+2,\dots \, , 
}
where $2t$ is the twist and $\eta^I_{nm,t\ell}$ is the anomalous dimension
which is perturbatively given as an expansion in $\lambda$, as in \expeta\
where $\eta_\ell\equiv \eta_{00,1\ell}$ with the index $I$ redundant. 
In general we have
\eqn\expert{
A_{nm}(u,v) = {p^2 \over N^2} \, \sum_{t,\ell}\,  u^t \sum_I a^I_{nm,t\ell} \, 
u^{{1\over 2} \eta^I_{nm,t\ell}} \, G^{(\ell)}_{2t + \ell + 4
+ \eta^I_{nm,t\ell}}\! (u,v) \, ,
}
where for $t = n+1,n+2, \dots$
\eqn\alim{
a^I_{nm,t\ell}\big |_{\eta^I_{nm,t\ell}\ne 0, \lambda\to 0} = a^I_{nm,t\ell,0}
\qquad {p^2 \over N^2} \sum_I a^I_{nm,t\ell,0} = {\hat A}_{nm,t\ell} \, , 
}
defines the zeroth order contribution of long supermultiplets in the conformal
partial wave expansion. Perturbation theory generates an expansion which
is expressible in the form
\eqn\Apert{
A_{nm}(u,v)_{\rm pert.} = {p^2 \over N^2} \, u^{n+1} \, {x \over x -\bx} 
\sum_{r=1,2,\dots } \!\! \lambda^r \sum_{s=0}^r  \,
\ln^s \! x\bx  \, \sum_{k=0}^\infty f_{nm,rs;k} (x) \, \bx^k  \, .
}
Comparing \Apert\ with \expert\ the $\ln x\bx$ terms determine the
anomalous scale dimensions $\eta^I_{nm,t\ell}$ as a series in $\lambda$.
For a given $k$ in the expansion in \Apert\ we must have $t\ge n+1+k$ in 
\expert. In appendix A it is shown how to define $f^{(j)}_{nm,rs}
= \sum_{k=0}^j \beta_{j,k} f_{nm,rs;k} $, $\beta_{j,j} =1$, so that 
$f^{(j)}_{nm,rs}(x)$ determines  the perturbative expansion $a^I_{nm,t\ell}, \,
\eta^I_{nm,t\ell}$ for just $t=n+1+j$. 

Here we consider for simplicity just the first order contributions 
to the anomalous dimensions which are constrained by
\eqn\expfm{\eqalign{
f^{(j)}_{nm,11}(x) = {}& \half x^j \!\! \sum_{\ell=-j-1}^\infty \sum_I  
a^{I}_{nm,n+j+1\,\ell,0} \, \eta^{I}_{nm,n+j+1\,\ell,1}\,  g_{n+j+3,\ell} (x) \, , \cr
f^{(j)}_{nm,22}(x) = {}& {\ts {1\over 8}} x^j \!\! \sum_{\ell=-j-1}^\infty \sum_I  
a^{I}_{nm,n+j+1\,\ell,0} \,
\big (\eta^{I}_{nm,n+j+1\,\ell,1}\big )^2  g_{n+j+3,\ell} (x) \, . \cr}
}
In \expfm\ we note that we may take $a^{I}_{nm,n+j+1\,\ell}=0$ for $\ell=-1$ and
contributions where $\ell<-1$, necessary for $j=1,2,\dots$, are in general
necessary but can be disregarded for the results obtained here. By using free
field results for $\sum_I a^{I}_{nm,n+j+1\,\ell,0}$ we are able to obtain from
the expansions \expfm\ results for $\langle \eta_{nm,t\ell,1} \rangle, \,
\langle \eta_{nm,t\ell,1}^{\, 2} \rangle$ where $t=n+j+1$. In general for unitarity
$\langle \eta_{nm,t\ell,1}^{\, 2} \rangle \ge \langle \eta_{nm,t\ell,1} \rangle^2$
with equality if just one operator is present.

We here apply this discussion to twist 4 operators for $p=3,4$ which,
as shown in section 4, are also decoupled in the large $N$ limit. At least
for these $p$ we find from  \pthree\ and \pfour\ using \onel\ and \Ftwo, with
\FIJ, we have a universal form to lowest order in an expansion in $u$,
\eqn\Atwo{\eqalign{
A_{1m}(u,v)_{\rm pert.} = {} & A_{1m,2}(u,v) + {\rm O}(u^3) \, , \quad m=0,1 \, , 
\cr
A_{00}(u,v)_{\rm pert.} = {}& A_{00,1}(u,v) + A_{00,2}(u,v) + {\rm O}(u^3) \, ,\cr}
}
where 
\eqnn\Atwoa
$$\eqalignno{
A_{00,2}(u,v) = {}& A_{11,2}(u,v) = {p^2 \over 6N^2}\, {u^2\over v^2} 
\big( - \half \lambda (1+v) \, \Phi^{(1)}(u,v) 
+ \quar \lambda^2 \K_{2,+}(u,v) +{\rm O}(\lambda^3) \big) \, , \cr
A_{10,2}(u,v) = {}& {p^2 \over 6N^2}\, {u^2\over v^2} \big(\half \lambda (1-v)\,
\Phi^{(1)}(u,v) - \quar \lambda^2 \K_{2,-} (u,v) +{\rm O}(\lambda^3) \big) \, ,
&\Atwoa \cr }
$$
for
\eqn\defK{
\K_{n,\pm}(u,v) = \quar (1\pm v^n) \Phi^{(1)}(u,v)^2  +
\Phi^{(2)}(u,v) \pm v^{n-2}\Phi^{(2)}(u/v,1/v) \, .
}
For $n=1,m=0,1$ we may restrict the expansion in \Apert\ to just $k=0$.
Using the results \Atoo\ with \athreeo\ or \Affour\ with \afouro\ for large
$N$ we have at zeroth order in $\lambda$,
\eqn\Atom{\eqalign{
\sum_I a^I_{10,2\ell,0} ={}& 
2^{\ell-1}\, {(\ell+2)!(\ell+3)!\over 3(2\ell+5)!} (\ell+1)(\ell+6) \, ,\cr 
\sum_I a^I_{11,2\ell,0} = {}& 2^{\ell-1} {(\ell+2)!(\ell+3)!\over 3(2\ell+5)!}
\big ( (\ell+3)(\ell+4)  + 2 \big ) \, .
\cr}
}
The leading terms in the expansion in \Apert\ are then given by
\eqn\flog{\eqalign{
f_{10,11;0}(x) = {}& {1\over 12} \, {1\over (1-x)^2} \ln(1-x) \, , \qquad
f_{11,11;0}(x) = - {1\over 12} \, {2-x\over x(1-x)^2} \ln(1-x) \, , \cr
f_{10,22;0}(x) = {}& - {1\over 48} \, {1\over x^2(1-x)^2} \big ( x^2 \!\Li_2(x)
+ \half x(3-x) \ln^2(1-x) \big )  \, , \cr
f_{11,22;0}(x) = {}& {1\over 48} \, {1\over x^2(1-x)^2} \big ( x(2-x) \Li_2(x)
+ \half (2-x+x^2) \ln^2(1-x) \big )  \, .  \cr}
}
Using the expansions of \flog\ obtained in appendix E in \expfm\ we have
\eqn\weight{\eqalign{
\l \eta_{11,2\ell,1} \r ={}& 2 \,{(\ell+3)(\ell+4) \big ( h(\ell+3) -1 \big)
\over (\ell+3)(\ell+4) + 2 } \, , \cr
\l \eta_{10,2\ell,1} \r ={}& 2\, {(\ell+2)(\ell+5) \big ( h(\ell+3) -1 \big)
-2 \over (\ell+1)(\ell+6)} \, , \cr}
}
and
\eqn\weightt{\eqalign{
\big ( (\ell+3)&(\ell+4) + 2 \big ) \l \eta_{11,2\ell,1}^{\, 2} \r \cr
&  = 4 (\ell+3)(\ell+4) \big ( h(\ell+3) -1 \big)^2 - 2 
\big ( h(\ell+3) -1 \big)\big ( h(\ell+3) +2 \big) \cr
\noalign{\vskip -5pt}
&\quad {} +2 \big ( (\ell+3)(\ell+4) -1 \big ) \sum_{r=2}^{\ell+3} 
{(-1)^r\over r^2} \, , \cr
(\ell+1)& (\ell+6) \l \eta_{10,2\ell,1}^{\, 2} \r \cr
&  =  4 (\ell+3)(\ell+4) \big ( h(\ell+3) -1 \big)^2 - 2 
\big ( h(\ell+3) -1 \big)\big ( h(\ell+3) +2 \big) \cr
\noalign{\vskip -5pt}
&\quad {} + 2 \big ( (\ell+3)(\ell+4) +1 \big ) \sum_{r=2}^{\ell+3} 
{(-1)^r\over r^2} - 6 \, . 
\cr}
}
It is evident that in general $\langle \eta_{1m,2\ell,1}^{\, 2} \rangle \ne
\langle \eta_{1m,2\ell,1}\rangle^2$ so that more than one operator must
contribute, unlike the twist two case. For both cases in \weightt\  
$\langle \eta^2 \rangle \ge \langle \eta \rangle^2$, for large $\ell$ 
$\langle \eta^2 \rangle - \langle \eta \rangle^2$ tends to $2-\pi^2/6$.
It is easy to check that when $\ell=0,1$
\eqn\lzero{
\l \eta_{11,20,1} \r = {\ts {10\over 7}} \, , \qquad
\l \eta_{10,21,1} \r = {\ts {5\over 2}} \, , \qquad
\l \eta_{11,20,1}^{\, 2} \r = {\ts {15\over 7}} \, , \qquad
\l \eta_{10,21,1}^{\, 2} \r = {\ts {25\over 4}} \, . \qquad
}
The results for $\langle \eta_{11,20,1} \rangle, \, 
\langle  \eta_{11,20,1}^{\, 2} \rangle$ are in accord with \hpert.
Since $\langle \eta_{10,21,1}^{\, 2} \rangle =
\langle \eta_{10,21,1} \rangle^2$ we expect that only one operator in this
case contributes in the large $N$ limit\foot{There is apparently one single trace
superconformal primary operator with $\ell=1$ and $\Delta=5$ \Heslop.}.

For twist 4 singlet operators we use formulae from appendix A to
determine
\eqn\flogz{\eqalign{
f^{(1)}_{00,11}(x) = {}& x f_{11,11;0}(x) - {1\over 2} \, {1\over x(1-x)} \, , \cr
f^{(1)}_{00,22}(x) = {}& x f_{11,22;0}(x) \cr
&{} + {1\over 16} \, {1\over x^3(1-x)} 
\big ((2-2x+x^2) \ln^2(1-x) + 2x(2-x) \ln (1-x) - 4 x^2 \big )  \, ,  \cr}
}
and from  \Atop\ or \Afop
\eqn\Atomz{
\sum_I a^I_{00,2\ell,0} =
2^{\ell-1}\, {(\ell+2)!(\ell+3)!\over 3 (2\ell+5)!}
\big ( (\ell+1)(\ell+6)+2 \big ) \, .
}
Using expansions from appendix E we obtain
\eqn\weightz{
\l \eta_{00,2\ell,1} \r = 2 \,{(\ell+3)(\ell+4) \big ( h(\ell+3) -1 \big) - 3 
\over (\ell+1)(\ell+6) + 2 } \, , 
}
and
\eqn\weighttz{\eqalign{
\big ( (\ell+1)&(\ell+6) + 2 \big ) \l \eta_{00,2\ell,1}^{\, 2} \r \cr
&  = 4 (\ell+3)(\ell+4) \big ( h(\ell+3) -1 \big)^2 + 2
\big ( h(\ell+3) -1 \big)\big ( 5h(\ell+3) - 2 \big) \cr
\noalign{\vskip -5pt}
&\quad {} +2 \big ( (\ell+3)(\ell+4) +5 \big ) \sum_{r=2}^{\ell+3}
{(-1)^r\over r^2} -24 + {12 \over (\ell+3)(\ell+4)} \, .
\cr}
}
For $ \ell = 0$ we have
\eqn\lzeros{
\l \eta_{00,20,1} \r = {\ts {7\over 4}} \, , \qquad\quad
\l \eta_{00,20,1}^{\, 2} \r = {\ts {27\over 8}} \, ,
}
in agreement with \hpert.

For completeness we also consider twist 6 operators where we may use the
perturbative results for $p=4$ correlation functions to obtain results
for anomalous dimensions which are not suppressed for large $N$. In this
case we may extend \Atwo\ to
\eqn\Athrees{\eqalign{
A_{2m}(u,v)_{\rm{pert.}}
= {}&  A_{2m,3}(u,v)+{\rm O}(u^4)\,,\qquad\qquad\qquad\quad m=0,1,2\,,\cr
A_{1m}(u,v)_{\rm{pert.}}
= {}& A_{1m,2}(u,v)+A_{1m,3}(u,v)+ {\rm O}(u^4)\,,\quad m=0,1\,,\cr
A_{00}(u,v)_{\rm{pert.}}
= {}& A_{00,1}(u,v)+A_{00,2}(u,v)+A_{00,3}(u,v)+ {\rm O}(u^4)\,, \cr}
}
where we use \Aone\ and \Atwoa\ and from \pfour, \FIJ\ for $p=4$,
\eqnn\secondtot
$$\eqalignno{
A_{22,3}(u,v){}& ={4\over 15 N^2}{u^3\over v^3}\big(-{\ts{1\over2}}
\lambda(1+v+v^2)\Phi^{(1)}(u,v) +
{\ts{1\over 4}} \lambda^2\K_{3,+}(u,v)+ {\rm O}(\lambda^3)\big)\,,\cr
A_{21,3}(u,v){}& ={\ts{3\over 5}}A_{10,3}(u,v)=
{4\over 5 N^2}{u^3\over v^3}\big(
{\ts{1\over 2}}\lambda(1-v^2)\Phi^{(1)}(u,v) - 
{\ts{1\over 4}}\lambda^2\K_{3,-}(u,v)+ {\rm O}(\lambda^3)\big)\,,\cr
A_{20,3}(u,v){}& =
{8\over 15 N^2}{u^3\over v^3}\big(-{\ts{1\over
 2}}\lambda(1-{\ts{1\over 2}}v+v^2)\Phi^{(1)}(u,v) + 
{\ts{1\over 4}}\lambda^2\K_{3,+}(u,v)+ {\rm O}(\lambda^3)\big)\,,\cr
A_{11,3}(u,v){}& =
{8\over 5 N^2}{u^3\over v^3}\big(-{\ts{1\over
 2}}\lambda(1+{\ts{1\over 6}}v+v^2)\Phi^{(1)}(u,v) +
{\ts{1\over 4}}\lambda^2\K_{3,+}(u,v)+ {\rm O}(\lambda^3)\big)\,,\cr
A_{00,3}(u,v){}& =
{4\over 5 N^2}{u^3\over v^3}\big(-{\ts{1\over
 2}}\lambda(1+{\ts{1\over 3}}v+v^2)\Phi^{(1)}(u,v) +
{\ts{1\over 4}}\lambda^2\K_{3,+}(u,v)+ {\rm O}(\lambda^3)\big)\, .
& \secondtot \cr}
$$
To first order in $\lambda$ we easily find from \secondtot,
\eqn\rhama{\eqalign{
f_{22,11;0}(x)& = -{1\over 120} \, {1\over x(1-x)^3}\,(3-3x+x^2)\ln(1-x)\,,\cr
f_{20,11;0}(x)& = -{1\over 120}\, {1\over x (1-x)^3}\,(3-3 x+2x^2)\ln(1-x)\,,\cr
f_{21,11;0}(x)& = {1\over 40} {2-x\over (1-x)^3}\, \ln(1-x)\,, \cr}
}
and taking into account the other terms in \Athrees\ we have
\eqn\fogive{\eqalign{
f^{(1)}_{10,11}(x){}&  = {1\over 24}\,{2-x\over x(1-x)^3}\,(1-x+x^2)\ln(1-x) + 
{1\over 12}{1\over (1-x)^2} \,,\cr 
f^{(1)}_{11,11}(x){}& = - {1\over 20}{1\over (1-x)^3}\, (3-3x+x^2)\ln(1-x)
-{1\over 12}{2-x\over x(1-x)^2} \,, \cr
f^{(2)}_{00,11}(x){}& = - {1\over 40}{1\over x(1-x)^3}\, 
\big ( (1-x)^4 +1 \big)\ln(1-x) -{1\over 12}{2-x\over (1-x)^2} \, . \cr}
}
At zeroth order from \Ahft, \Ahff\ and \afouro
\eqn\Atoms{\eqalign{
\sum_I a^I_{22,3\ell,0} ={}& 
2^{\ell-4}\, {(\ell+4)!(\ell+5)!\over 15(2\ell+7)!} \,
\big((\ell+1)(\ell+8)+18\big) \, ,\cr 
\sum_I a^I_{21,3\ell,0} = {}& 2^{\ell-4} {(\ell+3)!(\ell+4)!\over 
5(2\ell+7)!}\,
(\ell+1)(\ell+8)\big((\ell+2)(\ell+7)+8\big) \, ,\cr
\sum_I a^I_{20,3\ell,0} = {}& 2^{\ell-3} {(\ell+3)!(\ell+4)!\over 
15 (2\ell+7)!}\, (\ell+1)(\ell+2)(\ell+7)(\ell+8)\, ,\cr
\sum_I a^I_{10,3\ell,0} ={}& 
2^{\ell-4}\, {(\ell+3)!(\ell+4)!\over 3(2\ell+7)!}\,
(\ell+1)(\ell+3)(\ell+6)(\ell+8) \, ,\cr 
\sum_I a^I_{11,3\ell,0} = {}& 2^{\ell-3} {(\ell+4)!(\ell+5)!\over 5 
(2\ell+7)!}\, \big((\ell+1)(\ell+8) +{\ts {4\over 3}} \big) \, , \cr
\sum_I a^I_{00,3\ell,0} = {}& 2^{\ell-4} {(\ell+4)!(\ell+5)!\over 5 
(2\ell+7)!} \,\big((\ell+1)(\ell+8) +{\ts {2\over 3}} \big) \, . \cr}
}
Hence we obtain using expansions from appendix E
\eqn\aveta{\eqalign{
\l \eta_{22,3\ell,1}\r={}& 2\, {(\ell+3)(\ell+6)\big (h(\ell+4)-
{\ts{3\over 2}} \big )\over (\ell+3)(\ell+6)+ 8 }\,,\cr
\l \eta_{21,3\ell,1}\r={}&
2\, {(\ell+4)(\ell+5)\big((\ell+2)(\ell+7)\big(h(\ell+4)-{\ts{3\over 2}})-2\big)
\over (\ell+1)(\ell+8)\big((\ell+2)(\ell+7)+8\big)}\,,\cr
\l\eta_{20,3\ell,1}\r={}& 2\, {(\ell+3)(\ell+6)\big(h(\ell+4)-
{\ts {3\over 2}}\big )-3 \over (\ell+1)(\ell+8)}\,,\cr
\l \eta_{10,3\ell,1}\r={}& 2\, {(\ell+4)(\ell+5)\big (h(\ell+4)-
{\ts{3\over 2}}\big )-4 \over (\ell+1)(\ell+8)}\,, \cr
\l \eta_{11,3\ell,1}\r={}&
2\, {(\ell+3)(\ell+6)\big (h(\ell+4)-{\ts{3\over 2}}\big )- {\ts {10\over 3}}
\over (\ell+1)(\ell+8)+{\ts {4\over 3}}}\,,\cr
\l \eta_{00,3\ell,1}\r={}&
2\, {\big ((\ell+3)(\ell+6)+4\big )\big (h(\ell+4)-{\ts{3\over 2}}\big )
- {\ts {14\over 3}}\over (\ell+1)(\ell+8)+{\ts {2\over 3}}}\,.\cr}
}
In each case the leading behaviour for large $\ell$ is the same. The
corresponding results to \aveta\ for $\langle \eta^2 \rangle$ may also
be obtained but we omit these here.

\newsec{Conclusion}

The results of this paper show that the contributions of long multiplets
with twist at most $2p-2$ as $ \lambda \to 0$ are absent from the
operator product expansion of two BPS operators belonging to the $[0,p,0]$
representation in the large $N$ limit. This was demonstrated by explicit
calculation for $p=2,3,4$ using the results obtained by the AdS/CFT
correspondence. Such multiplets correspond to string states and are 
expected to have anomalous dimensions proportional to $\sqrt \lambda$
for large $\lambda$. In perturbation theory the anomalous dimensions are
given by an expansion in $\lambda$ without any $1/N$ suppression. Except
for the leading twist two case the anomalous dimensions cannot be determined
completely from the known perturbative results for four point correlation 
functions. In the twist two case we were able to recover the results
of perturbative calculations. For twist four and greater higher order
results would be necessary depending on the number of superconformal
primary operators present for each $\ell$. If only two are present
with anomalous dimensions $\eta_1,\eta_2$ then we would have the
relations for each $r=0,1,2\dots$ \Bia, 
\eqn\releta{
\langle \eta^{r+2} \rangle - (\eta_1+\eta_2) \langle \eta^{r+1} \rangle
+ \eta_1\eta_2 \langle \eta^r \rangle = 0 \, ,
}
where $\langle 1 \rangle =1$. A solution for $\eta_1,\eta_2$ is possible using
the $r=0,1$ relations if $\langle \eta^3 \rangle$
is known in addition to $\langle \eta \rangle, \langle \eta^2 \rangle$.
Using only $r=0$ then \lzero\ agrees with 
$\eta_{1,2} = \half (5 \pm \sqrt 5)\lambda$
found in \Bia\ for the lowest dimension scalar operators in the $[0,2,0]$
representation and \lzeros\ is in accord with $\eta_{1,2} = 
\quar (13 \pm \sqrt{41}) \lambda$ obtained in \mixing\ for singlet operators
with zeroth order dimension 4.
For just two operators \releta\ requires the consistency relation,
\eqn\che{
\langle \eta^4 \rangle - \langle \eta^2 \rangle^2 =
{\big ( \langle \eta^3 \rangle - \langle \eta \rangle \langle \eta^2 \rangle
\big )^2 \over \langle \eta^2 \rangle - \langle \eta \rangle^2} \, .
}
Alternatively we may extend the operator product expansion analysis
to correlation functions for BPS operators with different $p$, although
such cases have not been calculated either perturbatively or in the
large $N$ limit.

\bigskip
\noindent
{\bf Acknowledgments}
\medskip
We are grateful to Gleb Arutyunov, Paul Heslop and Emeri Sokatchev for many
helpful conversations. 

\vfil\eject
\appendix{A}{Conformal Partial Wave Expansions}

We consider first the general problem of expanding a general function of $u,v$
in terms of conformal partial waves
\eqn\exph{
u^a F(u,v) =  \sum_{{\ell=0,1,\dots\atop j=0,1,\dots}} \!\! c_{j,\ell}\,
u^{a+j} G^{(\ell)}_{\ell+2a+2j}(u,v)\, ,
}
where $F(u,v)$ is assumed to have a power series expansion in $u,1-v$.
If $F$ satisfies
\eqn\rph{
F(u,v) = \pm {1\over v^a} F(u/v,1/v) \, ,
}
then from \Gsym\ we must require $\ell$ to be respectively even, odd in \exph.

To determine $c_{j,\ell}$ in \exph\ we use the explicit form for 
$G_\Delta^{(\ell)}$ which, with $x,\bx$ defined as
in  \defxx\ and with $g_{t,\ell}$ as in \gell, was obtained in \DO
\eqn\solG{
G^{(\ell)}_{\ell+2t}(u,v) = {1 \over x -\bx } \Big (
x g_{t,\ell}(x)\, F(t-1,t-1;2t-2;\bx) - x \leftrightarrow \bx \Big ) \,  .
}
This satisfies
\eqn\Gid{
G_\Delta^{(\ell)}(u,v) = - (\quar u)^{\ell+1} G_\Delta^{(-\ell-2)}(u,v)\, ,
}
so that we may require that the coefficients in \exph\ satisfy
\eqn\cid{
- (\quar)^{\ell+1} c_{j,\ell} = c_{j+\ell+1,-\ell-2} \, .
}
The analysis depends on considering an expansion of $F$ in
the form
\eqn\whatwematch{
F(u,v) = {x \over x-\bx} \sum_{k=0}^{\infty}F_k(x)\, \bx^k \, .
}
Using the power series expansions of $F(t-1,t-1;2t-2;\bx)$ and $g_{t,\ell}(\bx)$ 
we may then match powers of $\bar{x}$ in \exph\ to find 
\eqn\Fid{\eqalign{
F_k(x) = {}& \sum_{j=0}^k \alpha_{k,j} \, x^{j} \sum_{\ell=0}^\infty
c_{j,\ell} \, g_{a+j,\ell}(x)\cr
&{}  - \sum_{\ell=0}^\infty  \sum_{j=0}^{k-\ell-1}
\alpha_{k,j+\ell+1}\, (\quar)^{\ell+1} c_{j,\ell} \, x^{j+\ell+1} 
g_{a+j+\ell+1,-\ell-2}(x) \, . \cr}
}
where we have used the definition \gell\ of $g_{t,\ell}$ and
\eqn\alkn{
\alpha_{k,j} = {1\over (k-j)!}\, {\big ( (a+j-1)_{k-j}\big )^2 \over
(2a+2j-2)_{k-j}} \, .
}
With the aid of \cid\ this may be easily rewritten as
\eqn\Fidd{
F_k(x) = \sum_{j=0}^k \alpha_{k,j} \, x^{j}\!\! \sum_{\ell=-j-1}^\infty \!
c_{j,\ell} \, g_{a+j,\ell}(x)\, .
}
As shown by Lang and R\"uhl (see the second paper in \Lang) this may
be inverted giving
\eqn\Fsol{
F^{(j)}(x) \equiv \sum_{k=0}^j \beta_{j,k} \, F_k(x) = x^{j}\!\!\!\!
\sum_{\ell=-j-1}^\infty \! c_{j,\ell} \, g_{a+j,\ell}(x)\, ,
}
where $\sum_{k=l}^j \beta_{j,k}\,\alpha_{k,l} = \de_{jl}$ which is satisfied by
\eqn\bkn{
\beta_{j,k} = (-1)^{j-k}{1\over (j-k)!}\, {\big ( (a+k-1)_{j-k}\big )^2 \over
(2a+k+j-3)_{j-k}} \, .
}
The result \Fidd\ then reduces the problem of determining $c_{j,\ell}$ to
matching single variable expansions which is more straightforward.
The first few $F^{(j)}$ are given by
\eqn\HH{\eqalign{
F^{(0)}(x) =  {}& F_0(x) \, , \qquad F^{(1)}(x) = 
F_1(x)-\half (a-1)\, F_0(x) \, ,\cr
F^{(2)}(x) =  {}&  F_2(x)- \half a\, F_1(x)+
{a(a-1)^2\over 4(2a-1)}\, F_0(x) \, . \cr}
}

\appendix{B}{Expansion Coefficients for Free Fields}

The determination of the
coefficients in the conformal partial wave expansions of $\hf(x,y)$
and, for free field theory, $\H_0(u,v;\sigma,\tau)$ for the cases $p=2,3,4$ 
considered here may be reduced to combinations of various basic expansions which 
are listed in this appendix.

The expansion of $\hf$, as in \exf, can be obtained by considering
\eqn\exxn{
x^{n+1} = \sum_{\ell=n}^\infty p_{n,\ell} \, g_{0,\ell+1}(x) \, , \qquad
x'{}^{n+1} = \sum_{\ell=n}^\infty (-1)^{\ell+1} p_{n,\ell} \, g_{0,\ell+1}(x) \, .
}
Since, with the definition \gell, $g_{0,\ell+1}(x) = (-\half)^{\ell+1} 
\sum_{n=\ell}^\infty \alpha_{n,\ell} \, x^{n+1}$, where $\alpha_{n,\ell}$
is given by \alkn\ with $a=2$, we then have from \bkn
\eqn\pnl{
(-\half )^{\ell+1} p_{n,\ell} = \beta_{\ell,n} = (-1)^{n-\ell} \, 
{(\ell !)^2 \over (2\ell)!} \, {(\ell+n)! \over (n!)^2 (\ell-n)! } \, ,
}
using \bkn\ for this case.
For subsequent use we note that from \gell\ we have for any integer $t$
\eqn\relg{
(-\half x)^t g_{t,\ell}(x) = g_{0,t+\ell}(x) \, .
}

For the analysis of $\H_0(u,v;\sigma,\tau)$ we require expansions as
in \exph\ with $a=2$ again. In this case $\beta_{j,k}$ is by \pnl\ and
\Fsol\ reduces to $(-\half)^{j+2} x^2 F^{(j)}(x) = \sum_\ell c_{j,\ell}
\, g_{0,j+\ell+2}(x)$. For expansion of free field expressions it is then
sufficient to use just \exxn\ and \pnl.

For the leading terms in each of \kHtwo, \threefree\ and \fourfree\
we require for $n=0,1$ and $2$,
\eqn\uvn{
u^n  =  \!\!\! \sum_{{\ell=0,1,\dots\atop t=n,n+1,\dots}} \!\!\! 
a^{(n)}_{t\ell}\,u^t\,G^{(\ell)}_{\ell+2t+4}(u,v)\,,\qquad
{u^n\over v^{n+2}}  = \!\!\! \sum_{{\ell=0,1,\dots\atop t=n,n+1,\dots}} 
\!\!\!
(-1)^\ell a^{(n)}_{t\ell}\,u^t\,G^{(\ell)}_{\ell+2t+4}(u,v)\, .
}
For $F(u,v)=u^n$ the method of obtaining the expansion coefficients described 
in appendix A then gives
$a^{(n)}_{t\ell} = (-2)^{\ell} ( \beta_{t,n} \beta_{t+\ell+1,n+1} -
\beta_{t,n+1} \beta_{t+\ell+1,n})$ or
\eqn\auvn{\eqalign{
a^{(n)}_{t\ell} = 
2^{\ell}{((\ell+t+1)!)^2\, (t!)^2\over (2\ell+2t+2)!\, (2t)!} \, &
(\ell+1)(\ell+2t+2)  \cr 
&{}\times {(\ell+t+1+n)!\, (t+n)! \over (\ell+t+1-n)!
\, (t-n)!\, ((n+1)!n!)^2}\, . \cr}
}
More generally, the expansion coefficients $a^{(n)}_{t\ell}$ are 
sufficient to compute the expansion
coefficients for the contributions from disconnected graphs
in the free-field four-point function for any $p=n+2$.

The sub-leading terms in the large $N$ limit require a variety of other
results. To determine
\eqn\vs{
{1 \over v}=\sum_{{\ell=0,2,\dots\atop t=0,1,\dots}}b_{t\ell}
\,u^t\,G^{(\ell)}_{\ell+2t+4}(u,v)\,,
}
we use \whatwematch\ and \Fsol\ with 
$\sum_{k=0}^j \beta_{j,k} = (-1)^j \beta_{j,0}$
to obtain  with $F(u,v) = 1/v$, $x^2F^{(j)}(x) = - \beta_{j,0} (x' + (-1)^j x)$.
Hence $b_{t\ell} = (1+(-1)^\ell)(-1)^{t+1}2^\ell\beta_{t,0}\beta_{t+\ell+1,0}$,
giving for $\ell$ even
\eqn\bvs{
b_{t\ell}=2^{\ell+1}{((\ell+t+1)!)^2\, (t!)^2\over (2\ell+2t+2)!\, (2t)!}
(-1)^t \, .
}
This allows the expansion coefficients for the $a$ term in
\kHtwo\ and the $b$ term in \threefree\ to be readily obtained.

In a similar fashion for
\eqn\uvs{
{u^2\over v^2}=\sum_{{\ell=0,2,\dots\atop
t=2,3,\dots}}c_{t\ell}\,u^t\,G^{(\ell)}_{\ell+2t+4}(u,v)\,,
}
we have, using now also $\sum_{k=0}^j \beta_{j,k} k = - (-1)^j \beta_{j,1}
= (-1)^j j(j+1) \beta_{j,0}$,
for this case from \Fsol\ $x^2F^{(j)}(x) = (1-(-1)^j)\beta_{j,0} (x^2 - x'^2) -
\beta_{j,1} ( (-1)^j x^2 + x'^2 + (1+(-1)^j)(x+x'))$.
Hence
$c_{t\ell} = (1+(-1)^\ell)2^\ell\{(1-(-1)^t)\beta_{t,0} 
\beta_{t+\ell+1,1} - (1+(-1)^t) \beta_{t,1} \beta_{t+\ell+1,0}
- (-1)^t \beta_{t,1} \beta_{t+\ell+1,1}\}$ giving for $\ell$ even
\eqn\cuv{\eqalign{
c_{t\ell}=2^{\ell}{((\ell+t+1)!)^2\, (t!)^2\over 
(2\ell+2t+2)!\, (2t)!} & \big((1+(-1)^t)t(t+1)(\ell+t)(\ell+t+3)\cr
\noalign{\vskip -6pt}
& - (1-(-1)^t)(t-1)(t+2)(\ell+t+1)(\ell+t+2)\big)\, . \cr }
}
In conjunction with recurrence relations given below, the expansion
coefficients $c_{t\ell}$ are sufficient to determine those for the $a$
term in \threefree\ and the $b$ and $c$
terms in \fourfree.

Additionally for the remaining $a$ terms in  \fourfree\ we require the
expansions. 
\eqn\uvt{
{u^2\over v}= \! \sum_{{\ell=0,1,\dots\atop 
t=2,3,\dots}}d_{t\ell}\,u^t\,
G^{(\ell)}_{\ell+2t+4}(u,v)\,,\qquad {u^2\over v^3}
= \! \sum_{{\ell=0,1,\dots\atop t=2,3,\dots}}(-1)^\ell 
d_{t\ell}\,u^t\, G^{(\ell)}_{\ell+2t+4}(u,v)\, .
}
For $F(u,v)=u^2/v$ then $x^2F^{(j)}(x) = (\beta_{j,0}(1-(-1)^j) + \beta_{j,1}) x^3
- \beta_{j,2}(x^2 + x+ x')$. This gives $d_{t\ell} = (-2)^{\ell}
\{ (1-(-1)^t)\beta_{t,0} + \beta_{t,1} ) \beta_{t+\ell+1,2} - \beta_{t,2}
( \beta_{t+\ell+1,1} + ( 1+ (-1)^{t+\ell}) \beta_{t+\ell+1,0} ) \}$ so that
for $\ell$ even
\eqn\deven{\eqalign{
d_{t\ell}={}& 2^{\ell-3}{((\ell+t+1)!)^2\,(t!)^2\over 
(2\ell+2t+2)!\, (2t)!} \cr
&\! {} \times\Big( (1+(-1)^t)t(t+1)(\ell+t)(\ell+ t + 3)
\big( (\ell+1)(\ell+2t+2)+2 \big ) \cr
\noalign{\vskip -4pt}
&\ \  {}+ (1-(-1)^t) (t-1)(t+2) (\ell+t +1)(\ell+ t + 2) 
\big( (\ell+1)(\ell+2t+2)-2 \big ) \Big ) \, ,  \cr}
}
and for $\ell$ odd
\eqn\dodd{\eqalign{
d_{t\ell}={}& 2^{\ell-3}{((\ell+t+1)!)^2\, (t!)^2\over                  
(2\ell+2t+2)!\, (2t)!}\, (\ell+1)(\ell+2t+2)  \cr 
&\ \ {} \times\big( (1+(-1)^t)t(t+1)(\ell+t+1)(\ell+ t + 2) \cr 
\noalign{\vskip -2pt}
&\qquad \ {}+ (1-(-1)^t) (t-1)(t+2) (\ell+t)(\ell+ t + 3) \big ) \, .  \cr}
}

All the above results for expansion coefficients satisfy the identity \cid.

For other results we may make repeated use of the
following recurrence relations,
\eqn\recurrencerels{\eqalign{
{(v-1)}\,G_{\De}^{(\ell)}(u,v)= {} & {2}\, G_{\De-1}^{(\ell+1)}(u,v)
+\half  u \, G_{\De-1}^{(\ell-1)}(u,v)\cr
& + {\ts{1\over 8}}f_{\De+\ell}\, u \, G^{(\ell+1)}_{\De+1}(u,v)
+{\ts{1\over 32}}f_{\De-\ell-2}\,u^2G^{(\ell-1)}_{\De+1}(u,v)\,,\cr
{(v+1)}\,G_{\De}^{(\ell)}(u,v)= {} & {2}\,G^{(\ell)}_{\De-2}(u,v)
+{\half} f_{\De+\ell}\,  G^{(\ell+2)}_{\De}(u,v)+\half 
u \, G_{\De}^{(\ell)}(u,v)\cr
& +{\ts{1 \over 32}}f_{\De-\ell-2}\,u^2 G^{(\ell-2)}_{\De}(u,v)
+{\ts{1\over 128}} f_{\De+\ell}\,f_{\De-\ell-2}\,u^2
G^{(\ell)}_{\De+2}(u,v)\,,}
}
where $f_\lambda=\lambda^2/(\lambda^2-1)$.  

\appendix{C}{Results for $\oD$ Functions and Loop Integrals}

We here list some results for the two variable functions
$\oD_{\De_1 \De_2 \De_3 \De_4}(u,v)$ which arise from AdS/CFT
integrals for large $N$ and in terms of which the perturbative 
results may also be expressed. Many properties are known, see 
\refs{\Freed,\DO,\ADHS}, only the significant ones in the present context 
are listed here. With the definition \defs\ for $s$ they may in general be 
expressed as power series as follows
\eqn\expD{\eqalign{
\!\!\!\!\!\!\!\!\!\oD_{\De_1\De_2\De_3\De_4}(u,v) = {}& 
\Gamma(-s )\,
{\Gamma(\De_1)\Gamma(\De_2)\Gamma(\De_3+s)\Gamma(\De_4+s)\over 
\Gamma(\De_1+\De_2)}\cr
&{} \times  G(\De_2,\De_3+s,1+s,\De_1+\De_2;u,1-v)\cr
& {} + \Gamma(s)\,
{\Gamma(\De_1-s)\Gamma(\De_2-s)\Gamma(\De_3)
\Gamma(\De_4)\over \Gamma(\De_3+\De_4)} \cr
&{} \times u^{-s}\,
G(\De_2-s,\De_3,1-s,\De_3+\De_4;u,1-v) \, ,  \cr}
}
where
\eqn\Gfunc{
G(\alpha,\beta,\gamma,\de;x,y)=\sum_{m,n=0}^{\infty}
{(\de-\alpha)_m(\de-\beta)_m\over m!\, (\gamma)_m}
{(\alpha)_{m+n}(\beta)_{m+n}\over n!\,(\de)_{2m+n}}\, x^m y^n\, .
}
The result \expD\ clearly satisfies \relDD. The series is convergent in 
the neighbourhood of $u,1-v \sim 0$. For other limits we may use
the symmetry relations for transpositions of $\De_i$,
\eqn\Dsym{\eqalign{
\oD_{\Delta_1\, \Delta_2\, \Delta_3\, \Delta_4}(u,v) = {}&
v^{-\Delta_2}\oD_{\Delta_1\, \Delta_2\, \Delta_4\, \Delta_3}(u/v,1/v) \cr
={}& \oD_{\Delta_3\, \Delta_2\, \Delta_1\, \Delta_4}(v,u) \cr
={}& u^{-\Delta_2} \,
\oD_{\Delta_4\, \Delta_2\, \Delta_3\, \Delta_1}(1/u,v/u) \, , \cr}
}
where other cases may be obtained by using the identities \relDD\
or
\eqn\DDs{
\oD_{\Delta_1\, \Delta_2\, \Delta_3\, \Delta_4}(u,v) =
v^{{1\over 2}(\Delta_1+\Delta_4-\De_2-\De_3)} \,
\oD_{\Delta_2\, \Delta_1\, \Delta_4\, \Delta_3}(u,v) \, .
}
The first relation in \Dsym\ is responsible for \Athree, \Afour\ and 
\Afoura\ obeying \AA.
We may also note that, directly from the representation \expD,
\eqn\refl{
\oD_{\Delta_1\, \Delta_2\, \Delta_3\, \Delta_4}(u,v) =
\oD_{\Sigma{-\Delta_3}\,\Sigma{-\Delta_4}\,\Sigma{-\Delta_1}\,
\Sigma{-\Delta_2}}(u,v)  \, , \quad \Sigma = \half \sum_{i=1}^4 \De_i \, .
}

The singularities  present in the result \expD\ arising from
$\Gamma(-s)$ for $s=0,1,2,\dots$ are cancelled by corresponding terms in 
the second expression on the right hand side of
\expD\ but this leads to $\ln u$ terms. The full result is then given
by \deDD\ and \Dsing\ with
\eqnn\Dexp
$$\eqalignno{
\!\!\!\!\!\!\!\!\!\!\oD_{\De_1\De_2\De_3\De_4}(u,v)_{\rm reg.} = {}&
{(-1)^{s}\over s!}\,
{\Gamma(\De_1)\Gamma(\De_2)\Gamma(\De_3+s)\Gamma(\De_4+s)
\over\Gamma(\De_1+\De_2)}\cr
& {} \times \Big( -\ln u\,G(\De_2,\De_3+s,1+s,\De_1+\De_2;u,1-v)& \Dexp\cr
&\quad {} +\sum_{m,n=0}^{\infty}{(\De_1)_m(\De_4+s)_m\over
m!\, (s+1)_m}{(\De_2)_{m+n}(\De_3+s)_{m+n}\over
n!\, (\De_1+\De_2)_{2m+n}}\,g_{mn}u^m(1-v)^n\Big)  \, , & \cr}
$$
where
\eqn\anothGop{\eqalign{
\!\!\!\!\!\!\!\!\!\!\!g_{mn}={}&  \psi(m+1)+\psi(s+m+1)+2
\psi(\De_1+\De_2 +2m+n)\cr
&{} -\psi(\De_1+m)-\psi(\De_4+s+m)-\psi(\De_2+m+n)-\psi(\De_3+s+m+n)\,.
}}
We may also note that from \Dsing
\eqn\Dsinga{\eqalign{
& \oD_{\De_1 \De_2 \, \De \, \De{+1}}(u,v)_{\rm sing.} - 
\oD_{\De_1 \De_2 \, \De{+1}\,  \De}(u,v)_{\rm sing.}  \cr
&{} = - u^{-s}(1-v)  \,
{\Gamma(\De_1-s)\Gamma(\De_2-s+1)\Gamma(\De)\Gamma(\De+1)\over
\Gamma(2\De+2) } \cr
& \quad {}\times
\sum_{m=0}^{s-1} (-1)^m (s-m-1)! \,
{(\De_1-s)_m (\De_2-s+1)_m (\De)_m (\De+1)_m\over m! \,
(2\De+2)_{2m} } \cr
\noalign{\vskip -6pt}
&\quad\qquad \ {} \times u^m F(\De_2-s+1+m,\De+m+1;2\De+2m+2;1-v) \, .\cr}
}

If $\Delta_i=0$ the integral defining the $\oD$ function reduces to that
for a three point function which may be directly evaluated giving
\eqn\redD{
\oD_{\Delta_1\, \Delta_2\, \Delta_3\, \Delta_4}(u,v)\Delta_4
\big |_{\Delta_4 \to 0} = \Gamma(\Sigma-\De_1)\Gamma(\Sigma-\De_2)
\Gamma(\Sigma-\De_3) \, u^{\De_3-\Sigma} v^{\De_1-\Sigma} \, ,
}
with results for other $\Delta_i=0$ obtained from \Dsym.

The derivation of  \Athree\ depends on the identity
\eqn\idthree{
\oD_{2523}(u,v) + \oD_{2532}(u,v)  = - \oD_{1533}(u,v) + u^{-1}v^{-2} \, ,
}
whereas to obtain \Afour\ we make use of
\eqn\idfour{\eqalign{
\oD_{3634}(u,v) + \oD_{3643}(u,v) ={}& - \oD_{2644}(u,v) + \oD_{2633}(u,v) \, , \cr
\oD_{2624}(u,v) + \oD_{2642}(u,v) ={}& - 2\oD_{2633}(u,v) - \big ( \oD_{1634}(u,v)
+ \oD_{1643}(u,v) \big ) \cr
\noalign{\vskip -4pt} 
&{} + 2u^{-1}v^{-3}(1+v) \, , \cr
\oD_{3623}(u,v) + \oD_{3632}(u,v) ={}& - \oD_{2633}(u,v) + u^{-2}v^{-2} \, , \cr
\oD_{4624}(u,v) + \oD_{4642}(u,v) ={}& - 2\oD_{4633}(u,v)  + \oD_{2644}(u,v)
- 2 \oD_{2633}(u,v) + u^{-2}v^{-2} \, . \cr}
}
To obtain \Afoura\ we also use
\eqn\idfoura{\eqalign{
\oD_{2624}(u,v) - \oD_{2642}(u,v) ={}& - \oD_{1634}(u,v) + \oD_{1643}(u,v) 
- 2u^{-1}v^{-3}(1-v) \, , \cr
\oD_{4624}(u,v) - \oD_{4642}(u,v) ={}& \oD_{3623}(u,v) - \oD_{3632}(u,v) -
\oD_{3634}(u,v) + \oD_{3643}(u,v) \, . \cr}
}
The results in \Afour\ may be further simplified by using
\eqn\idfoure{
\oD_{1634}(u,v) + \oD_{1643}(u,v) = - u\, \oD_{1733}(u,v)_{\rm reg.} - 4 \, {1\over
v^3}
\ln u + f(v) \, ,
}
with $f(v)$ given by a series in $1-v$.

Besides appearing in the large $N$ expansion the $\oD$ functions may also
be used as a generating function for the loop integrals $\Phi^{(L)}$ which
appear in perturbation theory. If $\de = \gamma+1$ in \Gfunc, which corresponds
to $\sum_i \De_i = 4$ in \expD, we have with the definitions \defxx\ from \DO
\eqn\redGG{\eqalign{\!\!\!\!\!
G(\alpha,\beta,\gamma ,\gamma+1;u,1-v)
=  {1\over x-\bx}& \big (
x F(\alpha,\beta;\gamma+1;x) F(\alpha-1,\beta-1;\gamma-1;\bx)\cr
\noalign{\vskip -4pt}
&\ {}  - x \leftrightarrow \bx \big )\, . \cr}
}
Using this we have
\eqn\Done{
\oD_{1{+\de}\, 1 1 \, 1-{\de}} (u,v) = {\pi \over \sin \pi \de} \Big ( 
- G_\de(x,\bx) + u^{-\de} G_{-\de}(x,\bx) \Big ) \, ,
}
where
\eqn\defG{
G_\de(x,\bx) = {1\over 1+ \de} \, {1\over x-\bx} \big ( x \, F(1,1+\de;2+\de;x) 
- x \leftrightarrow \bx \big )\, . 
}
By using standard hypergeometric identities we may then obtain
\eqn\Gig{
G_\de(x,\bx) = - {1\over x\bx} \, G_{- \de}(1/x,1/\bx) + {\pi \over \sin \pi \de} \,
{(-x)^{-\de} - (-\bx)^{-\de} \over x-\bx} \, ,
}
which, used in \Done, gives
\eqn\trD{
\oD_{1{+\de}\, 1 1 \, 1-{\de}} (u,v) = u^{-1-\de}\, \oD_{1{+\de}\, 1 1 \, 1-{\de}} 
(1/u,v/u) \, ,
}
in accord with \Dsym\ and \relDD.

{}From the identity \defG\ it is straightforward to see that
\eqn\Gexp{
G_\de(x,\bx) = \sum_{r=0}^\infty (-\de)^r \phi_{r+1}(x,\bx) \, ,
}
with $\phi_n$ defined in terms of polylogarithms as in \phin. Using \Gexp\
in \Gig\ is equivalent to standard identities relating $\Li_n(x)$ and
$\Li_n(1/x)$. The result \Gexp\ leads to a corresponding expansion of 
$\oD_{1{+\de}\, 1 1 \, 1-{\de}}$ by virtue of \Done. This may be rearranged 
in terms of generalised loop integrals where the definitions in \polyL\ for $L=1,2$
are extended \refs{\UD,\Isa}, with $u,v$ as in \defxx, to
\eqn\PhiL{
{1\over v} {\hat \Phi}^{(L)}(x,\bx) = {1\over L!} \sum_{n=L}^{2L}
{(-1)^n n! \over (n-L)! \, (2L-n)!} \, \ln^{2L-n}\! u \, \phi_n (x,\bx) \, ,
}
and we let $\Phi^{(L)}(u/v,1/v)= {\hat \Phi}^{(L)}(x,\bx)$.
The expansion obtained by using \Gexp\ in \Done\ may then be
re-expressed in the form
\eqn\DexpL{
\oD_{1{+\de}\, 1 1 \, 1-{\de}} (u,v) = {\pi \over \sin \pi \de}
\sum_{L=1}^\infty h_L(u,\de) \, {1\over v} {\hat \Phi}^{(L)}(x,\bx) \, ,
}
where
\eqn\hell{
h_L(u,\de) = {(-1)^{L}L! \over \ln^{2L-1}\! u} \sum_{r=0}^{L-1} 
{(2L-2-r)! \over (L-1-r)!\, r!} \, (\de \ln u ) ^r \big 
( u^{-\de} - (-1)^r \big ) \, .
}
The functions $h_L$ are linearly independent and, despite appearances, 
regular for 
$u=1$, $h_L(1,\de)= {2(L!)^2\over (2L)!}\, \de^{2L-1}$. It is easy to see that
\eqn\hinv{
h_{L}(u,\de) = u^{-\de} h_L(1/u,\de) \, ,
}
so that \trD\ is equivalent \relPh\ to for each $L$. As a consequence
of \DexpL\ $\Phi^{(L)}$ for each $L$ may therefore be obtained in terms of an
appropriate limit of the expansion of $\oD_{1{+\de}\, 1 1 \, 1-{\de}}$ 
to ${\rm O}(\de^{2L-2})$.

For the purpose of discussing perturbative results in terms of the
operator product expansion as in section 5 we need to analyse
$\Phi^{(L)}(1/u,v/u)$ for small $u,1-v$ or equivalently 
${\hat \Phi}^{(L)}(x,\bx)$ in the neighbourhood of $x,\bx=1$.
For $L=1$ it is sufficient to use
$\Phi^{(1)}(1/u,v/u)/u = \Phi^{(1)}(u,v)$ but for higher $L$ an expression
for the non analytic piece involving $\ln u$ may be obtained from
\eqn\Duv{\eqalign{
\oD_{1{+\de}\, 1 1 \, 1-{\de}} (v,u) = {}& \oD_{1 1 \, 1{+\de} \, 1-{\de}} (u,v) 
\sim -  {\pi \de \over \sin \pi \de} \, \ln u \, G(1,1+\de,1,2;u,1-v) \cr
&{} = -  {\pi \over \sin \pi \de} \, \ln u \, {1\over x - \bx} \big (
(1-x)^{-\de} - (1-\bx)^{-\de} \big ) \, , \cr}
}
where we have used \Dsym, the $\ln u$ part of \Dexp\ and \redGG. Using
the expansion \DexpL\ to ${\rm O}(\de^2)$ we may then obtain
\eqn\Phiu{
{1\over u} \, \Phi^{(2)}(1/u,v/u)  \sim \half  \ln u \ \phi_1(x,\bx) \, 
\ln(1-x) \, \ln (1- \bx) \, ,
}
neglecting terms which are just a power series in $u,1-v$. For application
in section 4 we may then note that $\ln(1-x) \, \ln (1- \bx) = {\rm O}(u)$.
To determine a suitable expansion for the for the non $\ln u$ terms in 
$\Phi^{(L)}(1/u,v/u)$ we may start from \defG\ and use
\eqn\fnew{
{x \over 1+ \de} \, F(1,1+\de; 2 + \de; x) = x^{-\de} \big ( \ln (1-x)
+ f_\de (1-x) \big ) \, ,
}
where $f_\de$ is expressible as a power series so that
\eqn\fexp{\eqalign{
f_\de(x) - \psi(1+\de) + \psi(1) ={}& \sum_{r=1}^\infty {(1+\de)_r \over r! \, r} \, 
x'{}^{r} - \ln(1- x)  \cr
= {}& \sum_{r=1}^\infty {(-\de)_r \over r! \, r} \, x^r 
= \int_0^x \! {1\over u} \big ( (1-u)^\de - 1 \big ) \, \d u  \, . \cr}
}
Using  \defG\ in \Done\ we then get
\eqn\Dvu{
\oD_{1{+\de}\, 1 1 \, 1-{\de}} (v,u) = - {\pi \over \sin \pi \de} \,
{1\over x-\bx} \Big ( (1-x)^{-\de} \big ( \ln x\bx + f_\de(x) +
f_{-\de}( \bx) \big ) - x \leftrightarrow \bx \Big ) \, .
}
This may be used to obtain results for $\Phi^{(L)}$ from \DexpL\ if we expand $f_\de$ 
in the form
\eqn\fexd{
f_\de(x) = \sum_{r=1,2,\dots} \!\! (-\de)^r \big ( p_r(x) - \zeta(r+1) \big ) \, , 
}
where
\eqn\defpr{
p_r(x) = {1\over r!}\int_0^x \! {1\over u} \, \big ( {-\ln (1-u)} \big )^r\, \d u 
= (-1)^r \sum_{n=r}^\infty 
\S_n^{(r)} {1\over n! \, n} (-x)^{n} \, , \qquad p_1(x) = \Li_2(x) \, ,
}
where $\S_n^{(r)}$ is a Stirling number of the first kind. It is easy to check that
\eqn\fid{
p_r(x)  = (-1)^r p_r(x') - {\ts {1\over (r+1)!}} \big( {-\ln (1-x)} \big )^{r+1} \, .
}
For $\Phi^{(2)}$ there is then an alternative expression of the form
\eqnn\Phiut
$$\eqalignno{
{1\over u} \, \Phi^{(2)}(1/u,v/u)  = {}& \big ( \half  \ln u \ \phi_1(x,\bx) 
- 3 \phi_2(x,\bx) \big ) \ln(1-x) \, \ln (1- \bx) + \half \phi_2(x,\bx) \,
\ln^2v \cr
&{}+ 3 \phi_1(x,\bx ) \big ( 2\zeta(3) - p_2(x) - p_2(\bx ) \big ) + 6 \,
{p_3(x)-p_3(\bx) \over x - \bx }  \, , & \Phiut \cr}
$$
where the right hand side may be readily expanded in powers of $x,\bx$.

\appendix{D}{Expansion Coefficients for Large $N$ Calculations} 

We here demonstrate how the first few expansion
coefficients in \expo\ are given by \solto\ with \cellab.  
We start by proving the result \exF, namely
\eqn\hypident{
F(a,b;2b;x) =\sum_{\ell=0,2,\dots}r_{ab,\ell}\,(\half x)^\ell\,
F(\ell+a,\ell+a;2\ell+2a;x)\,,
}
for $r_{ab,\ell}$ as in \cellab.
The summation over $\ell$ may be performed to rewrite the right-hand side of
\hypident, using standard $\Gamma$-function identities, as
\eqn\equivsum{
\sum_{n=0}^{\infty}\,{1\over n!}{(a)_n{}^2\over
(2a)_n}\,S_{n}(a,b)
\,x^n\,,
}
where 
\eqn\fourfthree{\eqalign{
S_{n}(a,b)= {}& {}_4 F_3 \left({-{n\over
2},\,{1\over 2}-{n\over 2},\,a-{1\over 2},\,a-b\atop a+{n\over 
2},\,a+{n\over
2}+{1\over 2},\,b+{1\over 2} }\Bigg|1\right)\cr
&+{4 n (n-1)(a-b)\over (2a+n)(2a+n+1)(2b+1)}\,{}_4 
F_3\left({1-{n\over 2},\,{3\over 2}-{n\over 2},\,a+{1\over 2},\,
a-b+1\atop a+{n\over 2}+1,\, a+{n\over 2}+{3\over 2},\,b+{3\over 2} }
\Bigg|1\right)\,,\cr}
}
in terms of standard ${}_4F_3$ hypergeometric functions.
All that is required now is to prove that the latter equals
\eqn\whatwemustprove{
T_n(a,b)={(2a)_n (b)_n\over (a)_n (2b)_n}\,,
}
when $n\geq 0$ so that \hypident\ immediately follows.  
This appears to be a non-standard hypergeometric 
identity however we may prove it
(by a method similar to that used by Pfaff in 1797 to prove the
Pfaff-Saalsch\"utz ${}_3F_2$ identity) by first establishing the
following recurrence
relation, namely,
\eqn\pfaffrecur{
S_n(a,b)=S_{n-1}(a,b)+{2(n-1)(2a+1)(2a+3)(a-b)\over
(2a+n-1)(2a+n)(2a+n+1)(2b+1)}
S_{n-2}(a+2,b+1)\,.
}
To see this, it is perhaps helpful to note 
that $S_n(a,b)$ may be rewritten as,
\eqn\rewriter{
S_n(a,b)=\sum_{k=0}^{[n/2]}S_{n,k}(a,b)\,,\qquad
S_{n,k}(a,b)=\left(1+{4 k\over 2 a-1 }\right)
{(-n)_{2k}(a-{1\over 2})_k(a-b)_k
\over k!(2a+n)_{2k}(b+{1\over 2})_k}\,,
}
whereby we may readily verify that
\eqn\recurVnk{
S_{n,k}(a,b)=S_{n-1,k}(a,b)+{2(n-1)(2a+1)(2a+3)(a-b)\over
(2a+n-1)(2a+n)(2a+n+1)(2b+1)}
S_{n-2,k-1}(a+2,b+1)\,,
}
so that, summing the latter over $k$, \pfaffrecur\ follows.
 We may then note that  \pfaffrecur\ with $S_0(a,b)=S_1(a,b)=1$
uniquely
defines $S_n(a,b),\,n\geq 0$.
Thus, as $T_n(a,b)$ in \whatwemustprove\ satisfies the same
recurrence relation \pfaffrecur\ and  
$T_0(a,b)=T_1(a,b)=1$ then $S_n(a,b)=T_n(a,b)$ for $n\geq 0$, as required.

Turning now to the proof of \expo, \solto\ with \cellab, we make use
of the general discussion in appendix A where we take
\eqn\dph{
F(u,v) = F(a,b;2b;1-v) \, ,
}
so that the definition \whatwematch\ of $F_k(x)$ now becomes
\eqn\whatdef{
(x-\bar{x})F\big (a,b;2b;x + \bx(1-x) \big )=
x \sum_{k=0}^{\infty}F_k(x)\, \bar{x}^k\,,
}
and the first few $F_k(x)$ are
\eqn\firstfew{\eqalign{
F_0(x) = {} & F(a,b;2b;x)\,,\cr
F_1(x) = {} & (\half a \,x -1){1\over x}F(a,b;2b;x)+{a(a-2b)\over 4(2b+1)} \, x 
F(a+1,b+1;2b+2;x)\,, \cr
F_2(x) = {} & {\ts{1\over 4}}\,a\big((a-b+1)x-2\big){1\over x}F(a,b;2b;x)\cr
&+{a(a-2b)\over 8(2b+1)} \, \big((a+b+1)x-2(b+1)\big)F(a+1,b+1;2b+2;x)\,,}
}
where we have used the identity
\eqn\whatusedo{
(x-1){\gamma\over \alpha\beta}\,
{{\rm d}\over {\rm d} x} 
F(\alpha,\beta;\gamma;x)=-F(\alpha,\beta;\gamma;x)+
{(\gamma-\alpha)(\gamma-\beta)\over \gamma(\gamma+1)}\,x\,
F(\alpha+1,\beta+1;\gamma+2;x)\,,
}
in the Taylor expansion of the left-hand side of \whatdef.
With \firstfew, we may use \HH\ to obtain
\eqn\simpssps{\eqalign{
&  F^{(0)}(x) = F(a,b;2b;x) \, , \cr
&  F^{(1)}(x) = {a b (a-2b)\over
2(4b^2-1)}\,x\,F(a+1,b+1;2b+2;x)- {1\over x} \, F(a-1,b-1;2b-2;x)\,,\cr
& F^{(2)}(x) = a(a+1)(b+1){(a-2b-1)(a-2b)\over 16 (2b+1)^2(2b+3)}
\,x^2\,F(a+2,b+2;2b+4;x)\cr
& \qquad \qquad \quad -a\,{a(a-3b-1)+b\over 4(2a-1)(2b+1)}\, F(a,b;2b;x)\,,}
}
where we have used the identity
\eqn\identused{\eqalign{
& \left(1-{\gamma(\alpha+\beta-1)-2\alpha\beta\over
\gamma(\gamma-2)}\,x\,\right)
F(\alpha,\beta;\gamma;x)\cr
&= F(\alpha-1,\beta-1;\gamma-2;x)
+\alpha\beta {(\gamma-\alpha)(\gamma-\beta)\over
\gamma^2(\gamma^2-1)}\,x^2\,
F(\alpha+1,\beta+1;\gamma+2;x)\,.}
}
Using \Fidd\ with \simpssps\ along with the expansion \hypident\
we may now easily determine $c_{j,\ell}$ as in \solto\ for $j=1,2$. 
It is easy to check that $\quar c_{0,0} = - c_{1,-2}$ and
$\quar c_{1,0} = - c_{2,-2}$ in accord with \cid.

An alternative approach to finding these results is to consider an expansion
of the conformal partial waves $G_\De^{(\ell)}$. By separating \solG\ into
two separate terms we have in general
\eqn\Gdiv{
G_\De^{(\ell)}(u,v) = 
f_{\De,\ell}(u,1-v) - (\quar u)^{\ell+1} f_{\De,- \ell-2}(u,1-v) \,,
}
and we may then expand $f_{\De,\ell}(u,1-v)$ in powers of $u$, extending the
result \lowt. The first few terms are then
\eqnn\expf
$$\eqalignno{
\!\!\!\!\! f_{\De,\ell}(u,x) ={}& g_{t,\ell}(x) - \quar(\ell-1) \, ug_{t+1,\ell-2}(x)
+{(t+\ell)^2(2t+\ell)\over 4(2t+2l-1)(2t+2l+1)}\, ug_{t+1,\ell}(x)\cr
&{}+ {\ts{1\over 32}}(\ell-2)(\ell-3) \, u^2 g_{t+2,\ell-4}(x)\cr
&{} - {(\ell-1)(2t+\ell)\over 16(2t+2l-3)(2t+2l+1)}
\Big ((t+\ell-1)(t+\ell)-{t-1\over 2t-1} \Big ) \, u^2g_{t+2,\ell-2}(x)\cr
&{}+ {(t+\ell)^2(t+\ell+1)^2(2t+\ell+1)(2t+\ell+2)\over 
32(2t+2l-1)(2t+2l+1)^2(2t+2l+3)}\, u^2g_{t+2,\ell}(x) + {\rm O}(u^3)  \, . &\expf \cr }
$$
The general form is $f_{\De,\ell}(u,x) = \sum_{p=0,1,\dots, q=0,1,\dots p} c_{p,q}\,
u^p g_{t+p,\ell-2q}(x)$, in accord with a very different treatment in \Except.
Each term in such an expansion is in accord with the symmetry condition \Gsym\
as a consequence of $g_{t,\ell}(x)=(-1)^\ell (1-x)^{-t} g_{t,\ell}(x')$.
For low $\ell$ terms involving $g_{t,-n}(x)$ for $n=1,2,\dots$ cancel between
the two contributions in \Gdiv. For a special case a related expansion was given
in the first paper in \HMR.

\appendix{E}{Expansion Coefficients for Perturbative Calculations}

The results of section 5 depend on the expansion of various functions $f(x)$,
such as listed in \frs, in terms of $g_{t,\ell}(x)$ for appropriate $t$. We 
here describe some further details as to how these were obtained. By letting 
$x^t f(x) \to f(x)$ and using \relg\ 
we need only consider expansion involving $g_{0,\ell}(x)$. In consequence
we need to determine coefficients ${\tilde f}_\ell$ such that
\eqn\exf{
f(x) = \sum_{\ell=1}^\infty 2^{\ell+1} {(\ell!)^2 \over (2\ell)!} \, 
{\tilde f}_\ell \, g_{0,\ell+1}(x) \, .
}
Since $g_{0,\ell}(x)=(-1)^\ell g_{0,\ell}(x')$ the expansion of $f(x')$ is
is as in \exf\ with ${\tilde f}_\ell \to (-1)^{\ell+1} {\tilde f}_\ell$.
By expanding $f$ so that
\eqn\exfn{
f(x) = \sum_{n=1}^\infty f_n \, x^{n+1} \, ,
}
and using \exxn, \pnl\ we obtain
\eqn\extf{
{\tilde f}_\ell = \sum_{n=1}^\ell f_n \, (-1)^{n+1}{(\ell+n)! \over
(n!)^2(\ell-n)!} \, .
}
The determination of ${\tilde f}_\ell$ is aided by using from \extf\
\eqn\aid{
\sum_{\ell=1}^\infty {\tilde f}_\ell\, y^\ell = - {1\over 1-y} \sum_{n=1}^\infty
f_n \, {(2n)! \over (n!)^2} \, \bigg ({-y \over (1-y)^2} \bigg )^n \, ,
}
where the right hand side may be determined in cases of interest. If
\eqn\Fnn{
S_k(y) = - \sum_{n=1}^\infty {1\over n^k}
\, {(2n)! \over (n!)^2} \, \bigg ({-y \over (1-y)^2} \bigg )^n \, ,
}
then
\eqn\Fdiff{
{1-y\over 1+y} \, y{\d \over \d y} S_k(y) = S_{k-1}(y) \, ,
}
so that we may find
\eqn\Fzz{
S_0(y) = {2y\over 1+y} \, , \qquad S_1(y) = -2 \ln (1-y) \, , \qquad
S_2(y) = 2 \big ( \Li_2(y) + \ln^2(1-y) \big ) \, .
}
Expanding $S_k(y)/(1-y)$ then gives ${\tilde f}_\ell$ in \extf\ for $f_n=n^{-k}$.
In other cases we have resorted to matching expansions obtained through
algebraic calculations of large numbers of terms.

Many of the results can be obtained from certain basic summations. We
consider first
\eqn\eL{
(-x)^p  \ln (1-x) = \sum_{\ell=1}^\infty 2^{\ell+1} 
{(\ell!)^2 \over (2\ell)!} \, a^{(p)}_\ell \, g_{0,\ell+1}(x) \, ,
}
and
\eqn\eLL{
-(-x)^p  \Li_2 (x) = \sum_{\ell=1}^\infty 2^{\ell+1} 
{(\ell!)^2 \over (2\ell)!} \, b^{(p)}_\ell \, g_{0,\ell+1}(x) \, ,
}
where the coefficients $a^{(p)}_\ell, b^{(p)}_\ell$ are obtained by taking
$f^{(p)}_n \to (-1)^{p+1}/(n-p+1)$, $(-1)^{p+1}/(n-p+1)^2$ for $n\ge p$
respectively in \extf. We find
\eqn\ap{\eqalign{
a^{(1)}_\ell = {}& 2h(\ell) \, , \cr
a^{(2)}_\ell = {}& 2\ell(\ell+1) \big ( h(\ell) - 1 \big ) \, , \cr
a^{(3)}_\ell = {}& \half (\ell-1)\ell(\ell+1)(\ell+2) 
\big ( h(\ell) - {\ts{3\over 2}} \big ) \, , \cr}
}
and
\eqn\bp{\eqalign{
b^{(1)}_\ell = {}& 2h(\ell)^2 \, , \cr
b^{(2)}_\ell ={}& 2 \ell(\ell+1) \big ( h(\ell)-1 \big )^2 - 2 h(\ell) +
\ell(\ell+1) \, . \cr}
}

With these results and \frs\ we have
\eqn\Lx{
x^3 f_{11}(x) = \half \big ( x  \ln (1-x) -  x' \ln (1-x') \big ) 
= - \!\! \sum_{\ell=2,4,\dots} \!\! 2^{\ell+2}
{(\ell!)^2 \over (2\ell)!} \, h(\ell) \, g_{0,\ell+1}(x) \, , 
}
and
\eqn\LLx{
x^3 f_{22}(x) = - \quar  \big ( x  \Li_2 (x) - x' \Li_2 (x') \big ) 
= - \!\! \sum_{\ell=2,4\dots} \!\! 2^{\ell+1}
{(\ell!)^2 \over (2\ell)!} \, h(\ell)^2 \, g_{0,\ell+1}(x) \, .
}
\Lx\ and \LLx\ directly lead to the expansions for 
$f_{1,1}$ and $f_{2,2}$ given by \exone\ and \extwo.

Further relevant expansions are given by
\eqn\exLi{
(-x')^p \Li_2 (x) = \sum_{\ell=1}^\infty 2^{\ell+1} 
{(\ell!)^2 \over (2\ell)!} \, b'{}^{(p)}_{\! \ell} \, g_{0,\ell+1}(x) \, ,
}
where we find
\eqn\bpr{\eqalign{
b'{}^{(1)}_{\! \ell} 
= {}& (-1)^\ell 2 \sum_{r=1}^\ell (-1)^{r} {1\over r ^2} \, , \cr
b'{}^{(2)}_{\! \ell} ={}& (-1)^\ell \bigg ( 2\ell(\ell+1) \sum_{r=1}^\ell {(-1)^r
\over r^2} + 2 h(\ell) + 1 \bigg ) - 1 \, . \cr}
}
Hence
\eqn\exfoz{
x^3 f_{10}(x) = (x+x') \Li_2(x) = \sum_{\ell=1}^\infty 2^{\ell+1} 
{(\ell!)^2 \over (2\ell)!} \, \big ( b^{(1)}_\ell + 
b'{}^{(1)}_{\! \ell}\big ) \, g_{0,\ell+1}(x) \, .
}

In addition for application in \exonea\ we also need to consider
\eqn\fderiv{
\!\! \sum_{\ell=2,4,\dots} \!\!  2^{\ell+1} {(\ell!)^2 \over (2\ell)!} \, 
h(\ell) \, g'{\!}_{0,\ell+1} (x) = - \sum_{\ell=3}^\infty 2^{\ell+1} 
{(\ell!)^2 \over (2\ell)!} \, d_\ell \, g_{0,\ell+1} (x)\, .
}
Expanding in powers of $x$ gives
\eqn\dell{\eqalign{
d_\ell = {}& 2 \sum_{n=2}^\ell  
(-1)^{n+1}{(\ell+n)! \over(\ell-n)!} \cr
&{} \times \! \sum_{j=2,4,\dots} \!\! {2j+1 \over (n-j)!\, (j+n+1)!} \, h(j) 
\big ( h(n) - h(j) - h(j+n+1) + h(2j+1) \big )  \cr
={}& \cases{h(\ell)^2 + \sum_{r=1}^\ell {(-1)^r \over r^2} \, , 
&if $\ \ \ell =3,5,\dots \, $,\cr
3 h(\ell)^2 - 2 h(2\ell)h(\ell) - \sum_{r=1}^\ell {1 \over r^2}(1+(-1)^r) \, ,  
&if $\ \ \ell =2,4,\dots \, $.\cr} \cr} 
}
With \exab\ and \bpr\ we then find
\eqn\resd{
{\bar d}_\ell = d_\ell - \half \big ( b^{(1)}_\ell +  b'{}^{(1)}_{\! \ell} \big ) 
= \cases{0 \, , &if $\ \ \ell =3,5,\dots \, $,\cr
2 h(\ell)^2 - 2 h(2\ell)h(\ell) - \sum_{r=1}^\ell {1 \over r^2} \, ,
&if $\ \ \ell =2,4,\dots \, $,\cr}
}
which, with the aid of \exfoz, is used in obtaining $b_{\ell,1}= {\bar d}_{\ell+2}$
in \aone\ from \exonea. 

{}From the result \frs\ for $f_{21}$ we have
\eqn\exfth{\eqalign{
x^3f_{21}(x) = {}& {\ts {3\over 4}}(x+x') \big ( \Li_3(x) - \Li_3(x') \big ) \cr
&{} + \quar (x-x') \ln(1-x) \Li_2 (x) + \quar (x+x') \ln(1-x') \Li_2 (x') \, .\cr}
}
To obtain its expansion we first consider
\eqn\eLth{
x  \Li_3 (x)  = \sum_{\ell=1}^\infty 2^{\ell+1} {(\ell!)^2 \over (2\ell)!} \, 
c_\ell \, g_{0,\ell+1}(x) \, , \quad
- x' \Li_3 (x) = \sum_{\ell=1}^\infty 2^{\ell+1} {(\ell!)^2 \over (2\ell)!} \, 
c'{\!}_\ell \, g_{0,\ell+1}(x) \, ,
}
which gives
\eqn\exc{\eqalign{
c_\ell = {}& \sum_{n=1}^\ell {1\over n^3} \, (-1)^{n+1}{(\ell+n)! \over (n!)^2(\ell-n)!} =
{2\over 3}\bigg ( \sum_{r=1}^\ell {1\over r^3} + 2 h(\ell)^3 \bigg ) \, , \cr
c'{\!}_\ell = {}& \sum_{n=1}^\ell \sum_{r=1}^n {1\over r^3} \, 
(-1)^{n+1}{(\ell+n)! \over (n!)^2(\ell-n)!} = 2 
\sum_{r=1}^\ell (-1)^{\ell+r} \bigg( 2h(r) {1\over r^2} - {1\over r^3} \bigg ) \, . \cr}
}
For the rest of $f_{21}$ it is then sufficient to expand
\eqn\eLtw{\eqalign{
-x \ln(1-x) \Li_2 (x)  = {}& \sum_{\ell=1}^\infty 2^{\ell+1} {(\ell!)^2 \over (2\ell)!} \, e_\ell \,
g_{0,\ell+1}(x) \, , \cr
x'\ln(1-x) \Li_2 (x) = {}& \sum_{\ell=1}^\infty 2^{\ell+1} 
{(\ell!)^2 \over (2\ell)!} \, e'{\!}_\ell \, g_{0,\ell+1}(x) \, , \cr}
}
which leads to
\eqn\eell{\eqalign{
e_\ell = {}& 4 \bigg ( \sum_{r=1}^\ell {(-1)^r \over r^3} - 2 \sum_{r=1}^\ell 
{1\over r} \sum_{s=1}^r {(-1)^s \over s^2} - h(\ell)^3 \bigg ) \, , \cr
e'{\!}_\ell = {}& - 4 (-1)^\ell \bigg ( 
\sum_{r=1}^\ell {1 \over r^3}\big ( 1 + 2(-1)^r \big ) - 4 
\sum_{r=1}^\ell {1\over r} \sum_{s=1}^r {(-1)^s \over s^2} + 3
h(\ell) \sum_{r=1}^\ell {(-1)^r \over r^2}  \bigg ) \, . \cr}
}
Hence \exfth\ gives
\eqn\expftt{
x^3 f_{21}(x) = \sum_{\ell=1}^\infty 2^{\ell-1} {(\ell!)^2 \over (2\ell)!} \,
\Big ( \big (3(c_\ell-c'{\!}_\ell)-e'{\!}_\ell \big )\big (1+(-1)^\ell\big ) 
- e_\ell \big (1 - (-1)^\ell\big ) \Big ) \, g_{0,\ell+1}(x) \, .
}
It remains to calculate
\eqn\ffderiv{
\!\! \sum_{\ell=2,4,\dots} \!\!  2^{\ell+1} {(\ell!)^2 \over (2\ell)!} \, h(\ell)^2 \,
g'{\!}_{0,\ell+1} (x) = - \sum_{\ell=3}^\infty 2^{\ell+1}
{(\ell!)^2 \over (2\ell)!} \, f_\ell \, g_{0,\ell+1} (x)\, .
}
By using an expression similar to \dell\ we may find
\eqn\fell{
f_\ell 
= \cases{h(\ell)^3 + 2\sum_{r=1}^\ell {1\over r} \sum_{s=1}^r {(-1)^s \over s^2} 
- \sum_{r=1}^\ell {(-1)^r \over r^3} \, ,&if $\ \ \ell =3,5,\dots \, $,\cr
\noalign{\vskip 4pt}
{\textstyle 
3 h(\ell)^3 - 2 h(2\ell)h(\ell)^2 - 2h(\ell) \sum_{r=1}^\ell{1\over r^2}
\quad\qquad\atop
\textstyle - 2\sum_{r=1}^\ell {1\over r} \sum_{s=1}^r {(-1)^s \over s^2}
+ \sum_{r=1}^\ell {1 \over r^3}(1+(-1)^r)} \, ,
&if $\ \ \ell =2,4,\dots \, $.\cr} 
}
Using these results in \etatwo\ we have for odd $\ell$
\eqn\fodd{
f_\ell + \quar e_\ell = 0 \, ,
}
whereas for even $\ell$,
\eqn\feven{\eqalign{
{\bar f}_\ell = f_\ell - {\ts {3\over 4}}(c_\ell - c'{\!}_\ell)
+ \quar e'{\!}_\ell
= {}& 2 h(\ell)^3 - 2 h(2\ell)h(\ell)^2 - 
2h(\ell) \sum_{r=1}^\ell{1\over r^2} \cr 
\noalign{\vskip -8pt}
&{} - \sum_{r=1}^\ell {1\over r}\sum_{s=1}^r {(-1)^s \over s^2} - 
\half \sum_{r=1}^\ell {1 \over r^3}(1-(-1)^r) \, ,\cr}
}
so that from \etatwo\ we have $\eta_{\ell,2} = 2 ({\bar f}_{\ell+2}
- {\bar d}_{\ell+2} h(\ell+2) )$, in accord with \etares.

For the analysis of the twist 4 case we have from \flog\
\eqn\simf{\eqalign{
x^4 f_{11,11;0}(x) = {}& {\ts{1\over12}}
\big ( x^2 \ln (1-x) + x'{}^2 \ln (1-x') \big ) \, ,
\cr x^4 f_{10,11;0}(x) = {}& {\ts{1\over 12}} 
\big ( (x^2+2x) \ln (1-x) - (x'{}^2+2x') \ln (1-x')\big ) \, . \cr}
}
Hence from \ap\ we have
\eqn\eLf{\eqalign{
x^4 f_{11,11;0}(x) = {}&\!\! \sum_{\ell=1,3,\dots}\!\! 
2^{\ell}{(\ell!)^2 \over 3(2\ell)!} \,  b^{(2)}_\ell \, g_{0,\ell+1}(x) \, , \cr
x^4 f_{10,11;0}(x) = {}&\!\!\sum_{\ell=2,4,\dots}\!\!
2^{\ell}{(\ell!)^2 \over 3(2\ell)!}
\, \Big (  b^{(2)}_\ell - 4h(\ell) \Big ) \, g_{0,\ell+1}(x) \, . \cr}
}
At the next order \flog\ gives
\eqn\simft{\eqalign{
x^4 f_{11,22;0}(x) = {}& -{\ts {1\over 48}}\big ( (2x^2 + x'^2 + x +x')\Li_2
(x) +  (2x'{}^2 + x^2 + x+x') \Li_2 (x') \big ) \, , \cr
x^4 f_{10,22;0}(x) = {}& -{\ts {1\over 48}}\big ( (2x^2 - x'^2 + x +x') \Li_2
(x) - (2x'{}^2 -x^2 +x+x') \Li_2 (x') \big ) \, . \cr}
}
The results \ap, \bp\ and \exLi\ then give
\eqn\eLLfour{\eqalign{
x^4f_{11,22;0}(x) = {}& \sum_{\ell=3,5,\dots} \!\! 2^{\ell-2}
{(\ell!)^2 \over 3(2\ell)!} \, \Big (2b^{(2)}_\ell - b'{}^{(2)}_{\!\ell} - 
b^{(1)}_\ell - b'{}^{(1)}_{\!\ell} \Big ) \, 
g_{0,\ell+1}(x) \, , \cr
x^4f_{10,22;0}(x) ={}& \sum_{\ell=2,4\dots} \!\! 2^{\ell-2}
{(\ell!)^2 \over 3(2\ell)!} \, \Big (2b^{(2)}_\ell + b'{}^{(2)}_{\!\ell} - 
b^{(1)}_\ell - b'{}^{(1)}_{\!\ell} \Big ) \, g_{0,\ell+1}(x) \, . \cr}
}
For singlet operators we may use from \flogz
\eqn\singf{
x^3 f^{(1)}_{00,11}(x) = x^4 f_{11,11;0}(x) + \half ( x+x')
= \sum_{\ell=1,3,\dots} \!\! 2^{\ell+1}
{(\ell!)^2 \over (2\ell)!} \, \big ( {\ts{1\over 6}} b^{(2)}_\ell -1 \big ) \,
g_{0,\ell+1}(x) \, , 
}
and
\eqn\singf{\eqalign{
x^3 f^{(1)}_{00,22}(x) -{}& x^4 f_{11,22;0}(x) \cr
= {}& {\ts{1\over 8}} ( x+x'-2)
\big ( \Li_2(x) + \Li_2(x') \big ) + {\ts{1\over 8}}( x-x') \ln (1-x) + 
\quar(x +x') \cr
= {}& \!\! \sum_{\ell=1,3,\dots} \!\! 2^{\ell-1}
{(\ell!)^2 \over (2\ell)!} \, \Big ( b^{(1)}_\ell + b'{}^{(1)}_{\! \ell}  - 
2h(\ell) - 2 + {2\over\ell(\ell+1)} \Big ) \, g_{0,\ell+1}(x) \, . \cr}
}
To obtain \singf\ we use
\eqn\Liz{
-\Li_2(x) - \Li_2(x') = \half \ln^2(1-x) = \!\! \sum_{\ell=1,3,\dots} \!\! 2^{\ell+1}
{(\ell!)^2 \over (2\ell)!} \, {2\over\ell(\ell+1)}  \, g_{0,\ell+1}(x) \, .
}

{}From \rhama\ we have
\eqn\rhamat{\eqalign{
x^5 f_{22,11;0}(x)& = {\ts {1\over 120}} 
\big ( x^3\ln(1-x)-x'^3\ln(1-x') \big ) \, , \cr
x^5 f_{20,11;0}(x)& = {\ts {1\over 120}}
(2 x^3+3x^2+6x)\ln(1-x)-(2x'^3+3x'^2+6x')\ln(1-x') \big ) \,,\cr
x^5 f_{21,11;0}(x)& = {\ts {1\over 40}} \big ( (x^3+x^2)\ln(1-x)+
(x'^3+x'^2)\ln(1-x') \big ) \,,\cr} }
and from \fogive
\eqn\rewriting{\eqalign{
x^4 f_{10,11}^{(1)}(x){}& = {\ts{1\over 24}} \big (
x^3\ln(1-x)+ x'^3\ln(1-x') \big ) + {\ts{1\over 12}}(x^2+ 2x
+ x'^2+ 2x') \,,\cr 
x^4 f^{(1)}_{11,11}(x){}& = {\ts{1\over 20}}\big (
x^3\ln(1-x)- x'^3\ln(1-x') \big ) + {\ts{1\over 12}}(x^2 -x'^2) \, , \cr
x^3 f^{(2)}_{00,11}(x){}& = {\ts{1\over 40}}\big (
(x^3-x^2)\ln(1-x)- (x'^3-x'^2)\ln(1-x') \big ) + 
{\ts{1\over 12}}(x^2 -x'^2) \, . \cr} 
}
Using \eL\ and \ap\ the required expansions may be easily read off.

\listrefs
\bye